\documentclass[12pt]{iopart}
\usepackage[utf8]{inputenc}
\usepackage{amsmath}
\usepackage{iopams}
\usepackage[caption=false]{subfig}
\usepackage{graphicx}
\usepackage{multirow}
\usepackage{color}
\usepackage{mathtools}

\begin{document}

\title{The Laplace method for energy eigenvalue problems in quantum mechanics}
\author{Jeremy Canfield$^1$, Anna Galler$^2$,  and James K. Freericks$^1$}
\address{$^1$Department of Physics, Georgetown University, 37th and O Sts. NW, Washington, DC 20057, USA\\
$^2$Max Planck Institute for the Structure and Dynamics of Matter, Luruper Chaussee 149, 22761 Hamburg, Germany}
\ead{james.freericks@georgetown.edu}
\vspace{10pt}
\begin{indented}
\item[]\today
\end{indented}
\begin{abstract}
   Quantum mechanics has about a dozen exactly solvable potentials. Normally, the time-independent Schr\"odinger equation for them is solved by using a generalized series solution for the bound states (using the Fr\"obenius method) and then an analytic continuation for the continuum states (if present). In this work, we present an alternative way to solve these problems, based on the Laplace method. This technique uses a similar procedure for the bound states and for the continuum states. It was originally used by Schr\"odinger when he solved for the wavefunctions of hydrogen. Dirac advocated using this method too. We discuss why it is a powerful approach for graduate students to learn and describe how it can be employed to solve all problems whose wavefunctions are represented in terms of confluent hypergeometric functions.
\end{abstract}

\maketitle

\section{Introduction}

The preponderance of quantum-mechanics instruction is via the differential equation form of the Schr\"odinger equation in position space. The methodology used (Fr\"obenius method for the series solution of differential equations) is essentially the same for undergraduate and for graduate-level courses. Unfortunately, this method of solution for bound-state problems is rarely used outside of the quantum mechanics classroom; in particular, it is rarely used in modern research. It also cannot be used to solve continuum problems. This begs the question: Why do we focus so much effort on the series solution of differential equations if students will not use this approach later in their careers? 

One answer might be ``because there is no other way to approach these problems.''~ But that answer is wrong! In this work, we show how one can employ the Laplace method to solve these problems with contour integrals in the complex plane. We advocate that this is likely to be a good option to use with graduate students, particularly because it will teach them how to use complex analysis to solve physics problems and prepare them for further use of complex analysis in many-body physics and in quantum field theory. Furthermore, this approach naturally allows one to solve both bound-state problems and continuum problems with similar efforts. The conventional differential equation approaches often struggle with solving continuum problems. It may also provide useful ways to experiment with and visualize the bound-state and continuum wavefunctions using the same numerical methods for both types of solutions. It can also provide practice with evaluating contour integrals numerically.  This method does have restrictions of its own, though. It can only solve problems whose wavefunctions can be represented in terms of \textit{confluent} hypergeometric functions.

This approach is as old as quantum mechanics itself. It was  introduced by Schr\"odinger when he solved for the spectrum of hydrogen in his first wave mechanics paper in 1926~\cite{schroedinger_1926}; the method was based on Schlesinger's famous differential equations textbook~\cite{schlesinger}. In 1937, Dirac advocated for using this same technique~\cite{dirac}. Oddly, he did not include it in his quantum mechanics textbook~\cite{dirac_textbook} (which was revised to its third edition ten years later), so it never was broadly adopted by the physics community. Modern texts that use it include Landau and Lifshitz~\cite{landau_lifshitz}, where it is employed in the appendix to describe the properties of special functions and Konishi and Paffuti~\cite{konishi_paffuti}, where they employ it to solve the linear-potential problem, but not other problems solved by confluent hypergeometric functions. Capri~\cite{capri} also uses it, but not as a general method. He suggests using an integral form (that corresponds to the Laplace method solution) to determine an integral representation for the bound-state wavefunctions of the Coulomb problem, and then evaluates them by residues.

Many instructors may be concerned that students are not ready to use complex analysis and this new math will need to be taught along with the quantum mechanics. How can one do both? Fortunately, one can develop the required basics rather quickly through the analogy between Stokes' theorem and Cauchy's theorem via the Cauchy-Riemann equations, which are identical to the condition of a vanishing curl of a two-dimensional vector field. The residue theorem follows almost immediately by investigating contour integrals (for contours that encircle the origin) of powers of $z$ (both positive and negative). One also needs to focus on how to determine the phase of complex numbers, relative to a reference point, when the complex plane has branch cuts in it due to the presence of multivalued functions (typically arising from noninteger powers). Unlike teaching the Fr\"obenius method, and often having to repeat it many times for the different applications, the Laplace method needs to be treated once---all of the confluent hypergeometric function solutions then follow in a straightforward manner from one single ``template.'' But most importantly, students will continue to use the complex variables knowledge in future courses and in research. This is why the effort is worthy of the investment it requires.

The Laplace method for solving differential equations is summarized in our earlier paper on Schr\"odinger's first solution for hydrogen~\cite{our_paper}. We provide a briefer summary here of this method. The basic idea is to find an integrating factor for the differential equation and then solve it using a contour integral constructed from this integrating factor.

The Laplace method works for arbitrary order linear  differential equations that have constant coefficients or linear coefficients (in the dependent variable). These differential equations take the form given by
\begin{eqnarray}
    \label{eq:lin_diff}
    \sum_m(a_m+b_m\xi)\Phi^{(m)}(\xi)=0,
\end{eqnarray}
where the $(m)$ superscript denotes the $m$th derivative of the function $\Phi(\xi)$ with respect to $\xi$.
For quantum mechanics solutions, we concentrate on the $m=2$ case because the time-independent Schr\"odinger equation is a second-order differential equation.
The solution to Eq.~(\ref{eq:lin_diff}) is constructed by introducing integrating factors and is represented in the form
\begin{eqnarray}
    \label{eq:lap_trafo}
     \Phi(\xi)=\int_\gamma e^{\xi z}R(z)dz,
\end{eqnarray}
with the integral being over a properly chosen contour $\gamma$ in the complex $z$ plane. The function $R(z)$ is determined from the integrating factor mentioned above.  

The form of the ansatz for $\Phi(\xi)$ allows us to compute derivatives by differentiating under the integral sign, 
\begin{eqnarray}
    \Phi^{m}(\xi)=\int_\gamma e^{\xi z}z^mR(z)dz
\end{eqnarray}
for properly chosen contours, where this procedure is well defined. Plugging this representation into the differential equation yields
\begin{eqnarray}
    \label{eq:diff_coeff}
    \int_\gamma e^{\xi z}\sum_m(a_m+b_m\xi)z^mR(z)dz=0,
\end{eqnarray}
which motivates the definition of two polynomials
\begin{eqnarray}
    P(z)=\sum_ma_mz^m \hspace{2em} \mathrm{and}  \hspace{2em} 
    Q(z)=\sum_mb_mz^m
\end{eqnarray}
that convert the differential equation into
\begin{eqnarray}
    \label{eq:diff_1}
    \int_\gamma e^{\xi z}\left[P(z)+Q(z)\xi\right]R(z)dz=0.
\end{eqnarray}
If the integrand is the derivative of a complex-valued function that has the same value at the endpoints of the contour $\gamma$, then we can immediately obtain the solution to the differential equation. We require that $R(z)$ satisfies the following differential equation
\begin{eqnarray}
    \label{eq:cond_1}
    P(z)R(z)=\frac{d}{dz}\left[Q(z)R(z)\right],
\end{eqnarray}
and that the function $V(z)$ defined by
\begin{eqnarray}
    V(z)=Q(z)R(z)e^{\xi z}
\end{eqnarray}
has equal values at the endpoints of the contour (or is single valued for a closed contour). Note that more than one contour can satisfy these conditions; indeed, for an order-$m$ equation, we know that exactly $m$ linearly independent solutions are represented by $m$ inequivalent contours.

The function $R(z)$ is then found by integrating Eq.~(\ref{eq:cond_1}) after dividing it by  $Q(z)R(z)$ and recognizing the resulting logarithmic derivative.
Solving,  yields $R(z)$ as
\begin{eqnarray}
    R(z)=\frac{1}{Q(z)}\exp\left(\int^{z}\frac{P(z')}{Q(z')}dz'\right).
\end{eqnarray}
Armed with $R(z)$, we immediately find the solution of the original differential equation to be
\begin{eqnarray}
    \label{eq:laplace_final}
    \Phi(\xi)=\int_\gamma e^{\xi z}\frac{1}{Q(z)}\exp\left(\int^{z}\frac{P(z')}{Q(z')}dz'\right)dz,
\end{eqnarray}
where the contour $\gamma$ needs to be chosen so that the vanishing endpoint condition is fulfilled. In quantum mechanics, we require the solution to satisfy additional properties, such as being always finite, or being square-integrable.

Next, we apply these methods to solve for the bound states of the simple harmonic oscillator in one- two- and three-dimensions. We do the same for the Coulomb problem in two- and three dimensions, as well as solving the Morse potential. We do not consider the spherical harmonics problem, or the P\"oschl-Teller and Hulth\'en potentials, because all of these cases require full hypergeometric functions for the wavefunction. These can be solved by the closely related Laplace transform method, as discussed in the work of Tsaur and Wang~\cite{tsaur_wang}.

\section{Bound states with the Laplace method}\label{sec:discrete}

We start with the application of the Laplace method to the bound-state problem. As mentioned above, we examine the simple harmonic oscillator in one-, two- and three-dimensions; the one-dimensional case is treated twice because there are two different forms for the wavefunction ansatz that are commonly used. We also cover the inverse $r$ potential (Coulomb problem) in two- and three-dimensions. Finally, we discuss the Morse potential, which is not commonly discussed in (physics-based) quantum-mechanics classes.

\begin{table}[htb!]
\centering
 \begin{tabular}{|c| c| c|c|} 
 \hline
 \multirow{2}*{\textbf{Problem}} & \multirow{2}*{\textbf{Potential}}  & \textbf{Independent}&\textbf{Wavefunc.} \\&& \textbf{Variable}&\textbf{Form} \\ \hline
 
 1D SHO, & \multirow{2}*{$V = \frac{1}{2}\mu\omega^2x^2$} & \multirow{2}*{$\xi = \frac{\mu\omega}{\hbar}x^2$} & \multirow{2}*{$\Phi(\xi)$}\\ Even &&&\\ \hline
 
 1D SHO, & \multirow{2}*{$V = \frac{1}{2}\mu\omega^2x^2$} & \multirow{2}*{$\xi = \frac{\mu\omega}{\hbar}x^2$} & \multirow{2}*{$x\Phi(\xi)$} \\ Odd &&&\\ \hline
 
 2D& \multirow{2}*{$V = \frac{1}{2}\mu\omega^2\rho^2$} & \multirow{2}*{$\xi=\frac{\mu\omega}{\hbar}\rho^2$} & \multirow{2}*{$\rho^{|m|}\Phi(\xi)e^{im\phi}$}\\SHO&&&\\ \hline
 
 3D  & \multirow{2}*{$V=\frac{1}{2}\mu\omega^2r^2$}& \multirow{2}*{$ \xi=\frac{\mu\omega}{\hbar}r^2$} & \multirow{2}*{$r^l\Phi(\xi)Y_l^m(\theta,\phi)$} \\SHO&&&\\ \hline

 2D  &\multirow{2}*{$V=-\frac{e^2}{\rho}$}& \multirow{2}*{$\xi = \sqrt{\frac{-2\mu E}{\hbar^2}}\rho$}&\multirow{2}*{$\rho^{|m|}\Phi(\xi)e^{im\phi}$}  \\ Coulomb &&&\\ \hline
 
 3D  &\multirow{2}*{$V=-\frac{e^2}{r}$}& \multirow{2}*{$\xi = \sqrt{\frac{-2\mu E}{\hbar^2}}r$} & \multirow{2}*{$r^l\Phi(\xi)Y_l^m(\theta,\phi)$}  \\ Coulomb &&&\\ \hline
 
 Morse  & \multirow{2}*{$V= V_0\left(e^{-2a x}-2e^{-a x}\right)$} & \multirow{2}*{$\xi =  \frac{2\sqrt{2\mu V_0}}{a\hbar}e^{-a x}$} & \multirow{2}*{$ \xi^{\frac{\sqrt{-2\mu E}}{a\hbar}}\Phi(\xi)$} \\ Potential&&& \\
 \hline
 
 1D SHO& \multirow{2}*{$V = \frac{1}{2}\mu\omega^2x^2$} &\multirow{2}*{$\xi = \sqrt{\frac{\mu\omega}{\hbar}}x$}&  \multirow{2}*{$e^{-\frac{\mu\omega}{2\hbar}x^2}\Phi(\xi)$} \\Method 2&&&\\ \hline
 
 \end{tabular}
\caption{Quantum-mechanical potentials with bound states that are analyzed in this work. For each, we give the form of the potential, the general form of the wavefunction where $\Phi$ is the unknown part determined by using the Laplace method, and the form for the independent variable $\xi$ used for each problem. Notably, $0\leq\xi <\infty$ in all cases but the last (where $-\infty\le\xi\le\infty$). In the table, $m$ is the $z$-component angular momentum quantum number with eigenvalues $\hbar m$ and  $l$ is the total angular momentum quantum number with eigenvalue $\hbar^2 l(l+1)$. Moreover, $\mu$ is the mass of the (effective) particle, $\omega$ is the frequency of the oscillator, $E$ is the energy of the corresponding energy eigenstate, $a$ is a real constant with units of inverse length, and $V_0$ has units of energy.} 
\label{Tab:table1}
\end{table}

The potentials for these different problems are summarized in Table \ref{Tab:table1}, where we use $\mu$ to denote the mass of the particle (sometimes this is the effective mass of a two-body problem), $\omega$ is the oscillator frequency, $e$ is the magnitude of the charge of an electron and of a proton, $V_0>0$ is an energy scale for the Morse potential, and $a>0$ is an inverse length for the Morse potential. Since all of these potentials do not have a Schr\"odinger equation that is in the Laplace form, we must do two additional things to arrive at the Laplace form: first, we construct an ansatz for the wavefunction, and compute the differential equation for the unknown function in the wavefunction ansatz; and second, we use a dimensionless independent variable (which sometimes is related to the original variable by a change in functional form). These choices are also summarized in Table \ref{Tab:table1}. The wavefunction ansatz arises from a number of different strategies. For the one-dimensional simple harmonic oscillator, we have a different form for the even and the odd solutions, in higher dimensions, we separate out the angular and radial degrees of freedom---in two dimensions we use $\rho$ and $\phi$ for the polar coordinates, while in three-dimensions we use $r$ for the radial coordinate and $\theta$ (polar angle to the $z$-axis) and $\phi$ (azimuthal angle in the $x$-$y$-plane) for the angular coordinates. The radial functions also have a power-law behavior as they approach the origin included in the ansatz. The second method for the simple harmonic oscillator in one dimensions removes a Gaussian factor from the wavefunction. The independent variable (always dimensionless and denoted by $\xi$) is proportional to the square of the (radial) coordinate for the simple harmonic oscillators, is linear for the Coulomb problems and is exponential for the Morse potential. For the second way we treat the simple harmonic oscillator in one dimension it is linear.

\begin{table}[htb!]
\centering

 \begin{tabular}{|c| c| c|} 
 \hline
 \multirow{2}*{\textbf{Problem}} & \textbf{Laplace Form of the} &\multirow{2}*{$\alpha_{\pm} = \frac{\beta\lambda\pm\delta}{2\lambda}$} \\ & \textbf{Schr\"odinger Equation}& \\ \hline
 
 1D SHO,  & \multirow{2}*{$\xi\Phi'' + \frac{1}{2}\Phi' + \left(\frac{E}{2\hbar\omega}-\frac{1}{4}\xi\right)\Phi=0$}& \multirow{2}*{$\frac{1}{4}\pm\frac{E}{2\hbar\omega}$} \\ Even &&\\
 \hline
 
 1D SHO,    & \multirow{2}*{$\xi\Phi'' + \frac{3}{2}\Phi' + \left(\frac{E}{2\hbar\omega}-\frac{1}{4}\xi\right)\Phi=0$} &\multirow{2}*{$\frac{3}{4}\pm\frac{E}{2\hbar\omega}$} \\ Odd &&\\
 \hline
 
 2D & \multirow{2}*{$\xi\Phi''+(|m|+1)\Phi' + \left(\frac{ E}{2\hbar\omega}-\frac{1}{4}\xi\right)\Phi=0$}&\multirow{2}*{$\frac{2|m|+1}{2}\pm\frac{E}{2\hbar\omega}$} \\SHO&&\\
 \hline
 
 3D  & \multirow{2}*{$\xi \Phi''+(l+\frac{3}{2})\Phi' + \left(\frac{ E}{2\hbar\omega}-\frac{1}{4}\xi\right)\Phi=0$}& \multirow{2}*{$\frac{2l+3}{4}\pm\frac{E}{2\hbar\omega}$} \\SHO&&\\
 \hline
 
   2D   & \multirow{2}*{$\xi\Phi''+(2|m|+1)\Phi'+\left(\frac{2\hbar}{a_0\sqrt{-2\mu E}} - \xi\right)\Phi=0$} & \multirow{2}*{$\frac{2|m|+1}{2}\pm\frac{\hbar}{a_0\sqrt{-2\mu E}}$}\\Coulomb&&\\
 \hline
 
 3D   & \multirow{2}*{$\xi\Phi''+2(l+1)\Phi'+\left(\frac{2\hbar}{a_0\sqrt{-2\mu E}} - \xi\right)\Phi=0$}& \multirow{2}*{$l+1\pm\frac{\hbar}{a_0\sqrt{-2\mu E}}$} \\Coulomb&&\\
 \hline
 
 Morse & \multirow{2}*{$\xi\Phi''+\left(2\frac{\sqrt{-2\mu E}}{a\hbar}+1\right)\Phi' + \left(\frac{\sqrt{2\mu V_0}}{a\hbar}-\frac{1}{4}\xi\right)\Phi=0$}&\multirow{2}*{$\frac{\sqrt{-2\mu E}\pm\sqrt{2\mu V_0}}{a\hbar}+\frac{1}{2}$} \\ Potential&& \\
 \hline
 
 1D SHO  & \multirow{2}*{$\Phi'' - 2\xi\Phi' + \left(\frac{2 E}{\hbar\omega}-1\right)\Phi=0$} &\multirow{2}*{\textbf{N/A}} \\ Method 2&&\\
 \hline
 
\end{tabular}
\caption{More details on the solutions for the wavefunctions using the Laplace method. The 1D SHO can be solved in two ways with the Laplace method, and we treat them both. In the second column, we provide the form of the Schr\"odinger equation obtained by the substitutions detailed in Table~\ref{Tab:table1}. Notably, for all but the last case, the coefficient of the first derivative term is either an integer or a half-odd integer, except in the Morse potential. It is larger than 1 for all of these cases except the first row.  Moreover, $a_0$ is the reduced Bohr radius ($a_0=\hbar^2/\mu e^2$).}  
\label{Tab:table2}
\end{table}

Note that the ansatz we use is not the standard one used in textbooks, where one incorporates the asymptotic behavior for large argument of the wavefunction as well. Instead, we are following the methodology of Schr\"odinger in his original paper, where only the behavior at the origin is incorporated into the ansatz.

Plugging the wavefunction ansatz and the functional form of the independent variable into the Schr\"odinger equation, yields a differential equation in the Laplace form for all of the examples we are working on in this paper. The results of this exercise are summarized in Table~\ref{Tab:table2}. There, you can see that all but the last row (which will be treated separately) have Schr\"odinger equations that have been transformed into the form: 
\begin{equation}\label{eq:laguerre_diffeq_form}
\xi\Phi''(\xi) + \beta \Phi'(\xi) +\left(\delta-\lambda^2\xi\right)\Phi(\xi) = 0,
\end{equation}
where $\beta,~\delta,~\lambda\in\mathbb{R}$ with $\lambda=1~\text{or}~\frac{1}{2}>0$ in the bound-state cases we consider here (when we later consider continuum solutions, the transformed Schr\"odinger equation for $\Phi$ has the same form, but $\lambda$ becomes an imaginary number, as we discuss below).  Notably, the form of Eq.~(\ref{eq:laguerre_diffeq_form}) is the same form used to treat the three dimensional hydrogen atom, as discussed in Refs.~\cite{schroedinger_1926,our_paper}. Nevertheless, the differences in the parameters in the Laplace form of the final differential equation, along with the different wavefunction ansatzes, mean that the analysis can be slightly different for some of these cases. We summarize the procedure in as general terms as possible, and show where the cases vary, as needed. Note further that this is not the standard Kummer equation, which arises when one includes the asymptotic behavior of the wavefunction for large argument as well. The form we use is in many respects easier to derive and simpler to work with. Of course, we will see the final answers are the same as one obtains with the standard methodology, as must be so.

As an illustration of this procedure, consider the even solutions of the one-dimensional simple-harmonic oscillator, as given in the first row. We use $\xi=\mu\omega x^2/\hbar$, so the kinetic energy operator becomes
\begin{equation}
    -\frac{\hbar^2}{2\mu}\frac{d^2}{dx^2}\to -2\hbar\omega\left (\xi\frac{d^2}{d\xi^2}+\frac{1}{2}\frac{d}{d\xi}\right )
\end{equation}
and the time-independent Schr\"odinger equation transforms into the equation in the top row of Table~\ref{Tab:table2} after we divide both sides by $-2\hbar\omega$ and we put all terms on the left hand side of the equation. The other rows are derived similarly.

To begin the Laplace method, we construct the $P$ and $Q$ polynomials as described in the introduction. They become
\begin{eqnarray}
    P(z) = \beta z+\delta,\\ Q(z) = z^2 -\lambda^2 = (z-\lambda)(z+\lambda).
\end{eqnarray} 
The ratio of $P/Q$ is then
\begin{equation}
    \frac{P(z)}{Q(z)} = \frac{\alpha_{+}}{z-\lambda}+\frac{\alpha_{-}}{z+\lambda},
\end{equation}
where \begin{equation}
\alpha_{+} = \frac{\beta\lambda +\delta}{2\lambda}~~\text{and}~~\alpha_{-} = \frac{\beta\lambda-\delta}{2\lambda}.
\end{equation}
This can be immediately integrated and exponentiated to form the integrand for the contour-integral form of the solution, which, up to a constant prefactor, is given by 
\begin{equation}\label{eq:gen_laguerre_diffeq}
    \Phi(\xi) = \int_\gamma dz~ e^{\xi z}(z-\lambda)^{\alpha_{+}-1}(z+\lambda)^{\alpha_{-}-1},
\end{equation}
where we still need to choose the contour $\gamma$. We must choose the contour $\gamma$ so that the value of the integrand of Eq.~(\ref{eq:gen_laguerre_diffeq}) multiplied by $Q(z)=(z-\lambda)(z+\lambda)$, has equal values at the endpoints. Closed contours will always satisfy this. If the contour is not closed, then the function that must be equal at the endpoints is 
\begin{equation}\label{eq:contour_condition_laguerre}
    V(z) = e^{\xi z}Q(z)R(z) = e^{\xi z}(z-\lambda)^{\alpha_{+}}(z+\lambda)^{\alpha_{-}}
\end{equation}
and this must hold for all $\xi$. Note that we always have $\alpha_+>0$ and $\alpha_++\alpha_-=\beta>0$, but $\alpha_-$ can take on positive or negative values.
In general, it is difficult to find contours where this function will be equal at the endpoints (for arbitrary values of $V$) if the path is not a closed path. However, special values, such as $0$ or $\infty$, are easier to enforce, because they automatically hold for all $\xi$; for $V=\infty$ one must use a careful limiting procedure to ensure the difference of $V$ at the two endpoints actually vanishes. 

The selection of these contours, and examining the integral over inequivalent contours is at the heart of the Laplace method. It requires a discussion of some technical details from complex analysis. Specifically the integrand is generically multivalued when $\alpha_\pm$ are nonintegers. Consider a complex valued function of the form $\mathcal{G}(z) = z^\alpha$. We can write this as 
\begin{equation}
    \mathcal{G}(z) = z^\alpha = e^{\alpha \log(z)}. 
\end{equation}
The complex logarithm is multivalued in the complex plane; consider 
\begin{equation}
\log(z) = \log\left(|z|e^{i\theta}\right) = \log(|z|) +i\theta.
\end{equation} 
Clearly, if we fix $|z|$ and move through $\theta$ from $0\to2\pi$, we will return to the same point on the complex plane, but the value of the logarithm will have picked up an additive shift by $2\pi i$ and thereby will not be a continuous (or even well-defined) function. So, we instead write the logarithm function as $\log(z) = \text{Log}(z)+2\pi i m,~m\in\mathbb{Z}$, where Log indicates the principal value of the logarithm function, which we take to be the complex logarithm function for $0< \theta<2\pi$ in this example; note that this choice is a domain that has a cut in the complex plane running from 0 to infinity along the positive real axis. $\text{Log}(z)$ is then single-valued, but the domain is smaller than the full complex plane due to the cut. This part of the complex plane given by $0< \theta < 2\pi$ is a \textit{branch} of the logarithm, defined to be a domain in the complex plane on which a function is single-valued. A branch is chosen to be maximal, in the sense that no point can be added to it and still maintain single-valuedness.

The boundary of a branch is called the \textit{branch cut}. From our definition of the complex logarithm, it is clear that the branch cut will be the ray where $\theta = 0$ (or more completely, where $\theta = (2n\pi$ for $n\in\mathbb{Z}$) originating at the origin; that is, the positive real axis plus the origin. In this case, the origin is a \textit{branch point} of the logarithm function since it is a point that is common to any branch cut one can draw for the logarithm. The branch cut can be any curve that extends from the origin to infinity to have a single-valued logarithm function.  Note that the complex plane, $\mathbb{C}$, does not include infinity, and as such there is no ``point at infinity;" instead approaching infinity, means that the complex number continues to increase without bound ($|z|\to\infty$). 

Critical to our work is a procedure that determines the phase of a number $z$ in a domain on the complex plane that has a branch cut. In the preceding discussion, the phase $\theta$ on the complex plane was defined in the standard convention: the phase is calculated with respect to the origin. That is, to determine the phase of $z$, we set the origin as the anchor of a vector of length $|z|$ starting from the horizontal (and oriented along the positive real axis), and rotate it counterclockwise until we reach the point $z$. Then, the phase of $z$ is the angle swept out by this line. When one needs the phase of a point $z$ relative to a reference point $(z_r)$, in a domain with a branch cut, one generalizes the process. We draw a path from the reference point to $z$ that does not cross any branch cut. Such a path will always exist for the cases we consider here. Then, we track the phase, relative to the reference point, as we move along the path, until we reach $z$. The phase is tracked by observing how the angle between the reference point and a line whose endpoint traverses along the path varies from the reference point to $z$; explicit examples of how this works will be given below.  Note that while the path (and hence the tip of the line) never crosses a branch cut, the arrow drawn from the reference point to the tip often does cross a branch cut. Hence, the final phase that is used, may look like we rotated an arrow from the reference point to the final point $z$ across a branch cut. But, as we see above, because we follow a path that does not cross the branch cut, this is fine. It is, however, a clear source of confusion for students, so this point must be emphasized during instruction. 

We now return to the complex power function, which becomes 
\begin{equation}
    \mathcal{G}(z) =z^\alpha =  e^{\alpha\text{Log}(z)}e^{2\pi i m\alpha}.
\end{equation}
Clearly, if $\alpha\in\mathbb{Z}$, then $\mathcal{G} = e^{\alpha\text{Log}(z)}=z^\alpha$, which is single-valued. Otherwise, it is multivalued. In general, $\alpha_{\pm}$ in Eq.~(\ref{eq:gen_laguerre_diffeq}) are not integers. This means that the integrand is not single-valued in general. We must draw branch cuts to restrict the complex plane to a domain where the integrand is single valued. The contour is then required not to cross any branch cut (but it can have endpoints at branch cuts or branch points). 

For a given multivalued function, there can be many inequivalent ways to draw branch cuts (the only requirement is that the integrand is single-valued within the domain that has the branch cuts drawn). From the form of the integrand in Eq.~(\ref{eq:gen_laguerre_diffeq}), we know that the branch points will be determined by the exponents $\alpha_{\pm}$; that is, $\pm\lambda$ is a branch point if and only if $\alpha_{\pm}\notin\mathbb{Z}$, respectively. By construction, $\alpha_{+}+\alpha_{-} = \beta$, so looking at Table~\ref{Tab:table2}, we immediately see two cases: those where $\beta\in\mathbb{Z}$ (first six rows) and those where $\beta\notin\mathbb{Z}$ (Morse potential). These two cases have different allowed branch cuts. In addition to determining the branch cuts and the allowed contours (which must have $V(z)$ be equal at the endpoints for all $\xi$ values), we also require the function $\Phi(\xi)$ to be nonsingular for all $\xi$. This condition is different from the conventional requirement of square-integrability of the wavefunction, but it is the condition needed for the Laplace method. We will find all bound-state wavefunctions that satisfy this condition are also square integrable, but we do not prove this.

First, we can eliminate any closed contour that does not cross a branch cut or contain any branch points. This is because such a closed contour always gives a zero integral due to Cauchy's theorem and the fact that the function is analytic inside the domain defined by the branch. An open contour that goes from a finite point $z$ to another finite point $z'$ inside the domain (that does not cross a branch cut) is also ruled out because it will not be possible to satisfy $V(z)=V(z')$ for all $0\le \xi <\infty$.

These restrictions then imply that contours must go to infinity, or end at a branch point (technically they could also end anywhere on a branch cut, but similar to the open contour case above, one won't be able to satisfy the condition $V(z)=V(z')$ for all allowed $\xi$ in that case). Rather than determine all possible contours next, we now consider the condition that the solution must be finite as $\xi\to\infty$.
Since there is a factor of $e^{\xi z}$ in the integrand, no contour can go to infinity with $\text{Re} (z)>0$ without the integral diverging as $\xi\to+\infty$. Moreover, the contour cannot have an end point at $z=\lambda$ because using the stationary phase method to asymptotically determine the limit of the integral for large $\xi$, again  has an exponential term that will diverge (for details see \cite{our_paper}).

\begin{figure}
    \centering
    \includegraphics[width=\columnwidth]{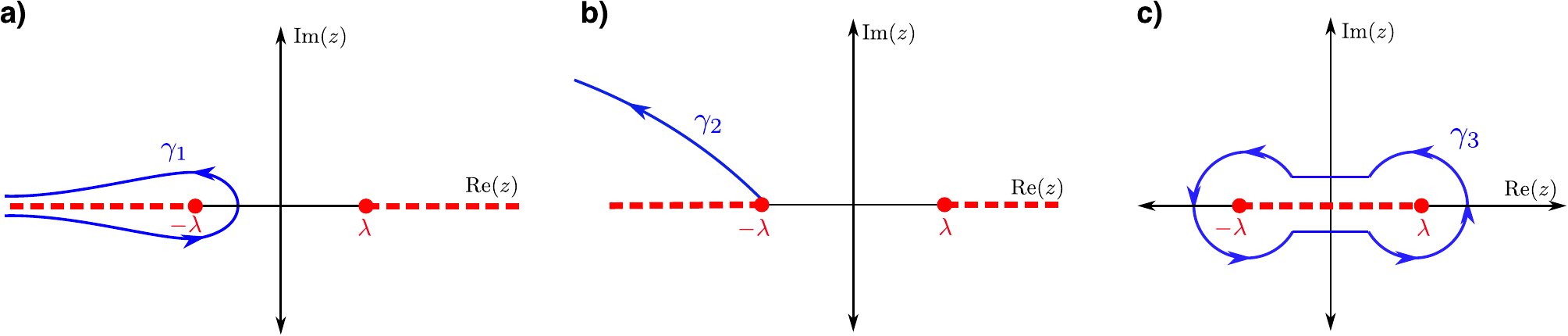}
    \caption{The contours we consider when $\alpha_{\pm}$ are not integers. The branch cuts in each case are shown by the dashed red lines.}
    \label{fig:cont_all}
\end{figure}

For all cases, we can have a branch cut running from $-\infty$ to $-\lambda$ along the negative real axis and from $\lambda$ to $\infty$ along the positive real axis. When $\beta$ is an integer, we can instead have the branch cut run along the real axis between $-\lambda$ and $\lambda$. These two possibilities are depicted in Fig.~\ref{fig:cont_all}.

We describe next the three possible contours we can have for all $0<\xi$ (we will determine the behavior for $\xi=0$ last). Two contours can be drawn when we have the branch cuts extending from $-\lambda$ to $-\infty$ along the negative real axis and from $\lambda$ to $\infty$ along the positive real axis. The first one, $\gamma_1$, is the so-called Hankel contour, which starts at $-\infty$ just below the real axis, loops around the branch point, and goes back to $-\infty$ above the real axis (see Fig.~\ref{fig:cont_all}~a). The second possibility, $\gamma_2$, starts at the branch point at $-\lambda$ and goes to $-\infty$ (either above or below the real axis, both end up giving the same asymptotic behavior as $\xi\to 0$; we draw it above the real axis here). The final contour can be drawn when the branch cut extends from $-\lambda$ to $\lambda$ (which is possible only because $\alpha_++\alpha_-=\beta$ is an integer). This contour, $\gamma_3$, encircles the branch cut and the branch points; it is sometimes called the dog-bone contour. Note that a similar $\gamma_2$ exists when the branch cut runs from $-\lambda$ to $\lambda$, but it behaves the same as the $\gamma_2$ that we analyze. A stationary phase analysis of $\gamma_3$ shows that it diverges as $\xi\to\infty$, so it is ruled out~\cite{our_paper}.

Now, we need to determine if any of the remaining contours yield a finite result as $\xi\to0$. First, in the case of the Hankel contour, the phase of the integrand will be different below and above the real axis, since we drew a branch cut along the negative real axis. Hence,  $\Phi(\xi)\to\infty$ for $\xi=0$ for the Hankel contour, because the integrand remains finite but the contour extends an infinite length, and the difference in the phases means those infinite contributions do not cancel. Thus, we can rule out the $\gamma_1$ contour.  For the contour $\gamma_2$ from $-\lambda\to\infty$ (in the left half plane), if $\alpha_-<0$, the integral will diverge as $z\to -\lambda$. If $\alpha_-\ge 0$, we find that the contour does satisfy the condition that $V(z)$ has equal values (in fact $V(z)=0$) at the endpoints of the contour $\gamma_2$ for $\xi>0$. But, when $\xi =0$, $V(-\lambda) = 0$ because there is a factor of $z+\lambda$ raised to a positive power (it is a finite number when $\alpha_-=0$), while for $z\to -\infty$, $|V|$ goes like $|z|^{\beta}\to\infty$. So, our condition that the endpoints have equal values of $V$ for all values of $\xi$ does not hold for $\xi=0$, eliminating $\gamma_2$. This eliminates all possible contours when $\alpha_\pm$ are not integers.

We still have to consider the case when $\beta\notin\mathbb{Z}$ for the Morse potential. This does not change any of the prior analysis; it simply restricts us to exclude the cases where the branch cut ran from $-\lambda$ to $\lambda$---hence, we have no dog-bone contour to consider.

Since there are no valid solutions for the general case,  we must re-evaluate our prior assumption, that in general $\alpha_{\pm}\notin\mathbb{Z}$, to see if we can find another contour in that case. It is clear that we will require at least one of them to be an integer, in general. When $\beta\in\mathbb{Z}$, there is no ambiguity, as in this case, if one is an integer, they both must be. When $\beta\notin\mathbb{Z}$, the analysis is more complex. Note that, as shown in Table~\ref{Tab:table2}, these cases are the one and three dimensional oscillators and the Morse potential, since $a\notin\mathbb{Z}$, in general. Recalling that in all cases, $\delta,\lambda>0$, we know that $\alpha_{+}>\alpha_{-}$. Moreover, the only way to obtain a new  contour is to have one that surrounds a pole, that is a point $z'$ at which $f(z)\propto (z-z')^{-N}$ for $N\in\mathbb{Z}^+$. Since $\alpha_++\alpha_-=\beta>0$ and $\alpha_+>\alpha_-$, we can only have $\alpha_{-}$ be a non-positive integer (we can have $\alpha_{-} = 0$,  because in Eq.~(\ref{eq:gen_laguerre_diffeq}), the exponent of $(z+\lambda)$ is $(\alpha_{-}-1)$, so $\alpha_{-}=0$ will result in a first-order simple pole). That is, 
\begin{equation}
|\alpha_-|=N,~N=0,1,2... 
\end{equation}
We will find that $\alpha_-=0$ corresponds to the ground state, and $\alpha_{-}<0\in\mathbb{Z}$ will yield higher order poles, corresponding to excited states.

\begin{figure}
    \centering
    \includegraphics[width=0.8\columnwidth]{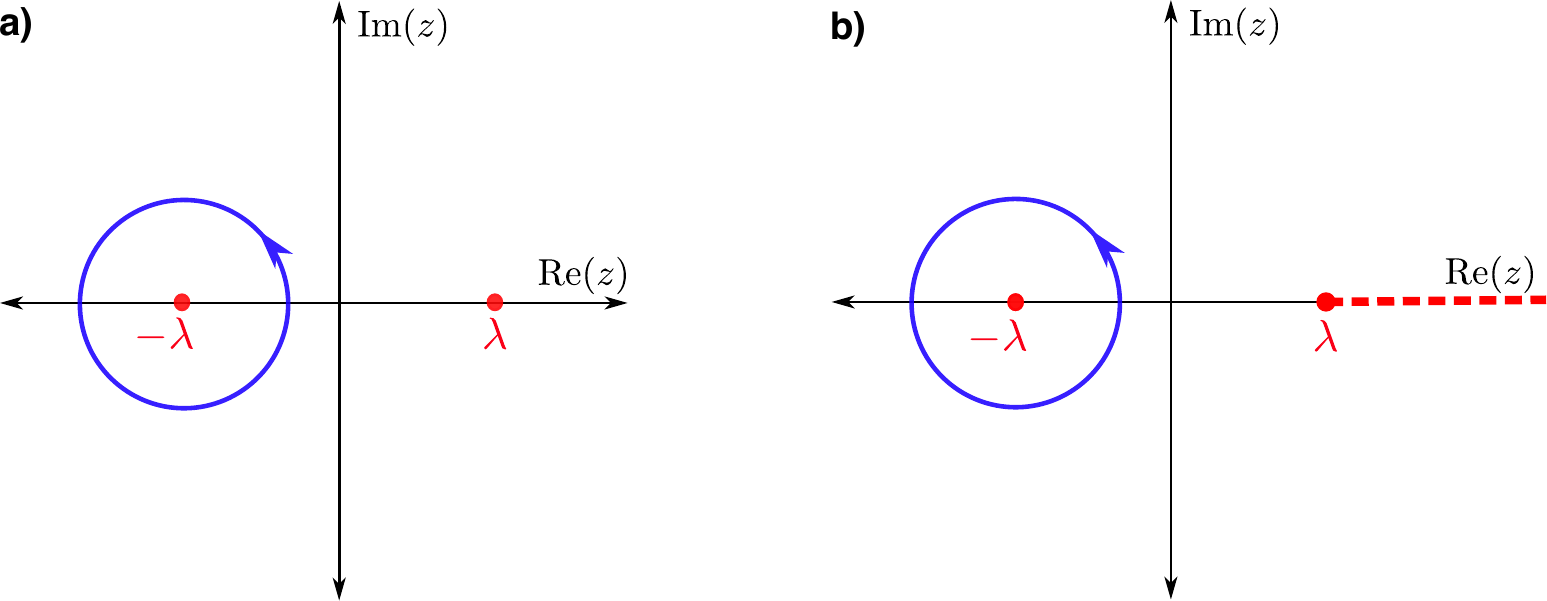}
    \caption{Contours that lead to the correct solution for the wavefunction and provide the quantization condition for the energy. The contour is always a closed contour in the counter-clockwise direction encirling the point $z=-\lambda$. In some cases, shown in panel (b), a branch point remains at $z=\lambda$ and the branch cut runs from there to $z=\infty$ along the real axis.}
    \label{fig:cont_solns1}
\end{figure}

Note that the constraint that $\alpha_{-}$ is a nonpositive integer yields the quantization condition for each bound state  (see Table~\ref{Tab:table3}). Key to our analysis is that integer exponents lead to single-valued powers, so $z=-\lambda$ is no longer a branch point, and we can enclose it with a closed contour (Fig.~\ref{fig:cont_solns1}). Since this point is a pole of order $|\alpha_{-}|+1$, this contour is non-trivial and distinct from the previously analyzed ones. Notably, the cases when $\beta\notin\mathbb{Z}$ are only slightly different; $z=\lambda$ is still a branch point, and we choose the branch cut to run from $\lambda\to\infty$ along the positive real axis, as shown in Fig.~\ref{fig:cont_solns1}~(b). Thus, all cases of the form in Eq.~(\ref{eq:gen_laguerre_diffeq}) lead us to the same new contour, a closed contour about a pole of order $N+1$ at $z=-\lambda$. This yields the solution in the form of a closed contour integral (with the contour encircling the point $z=-\lambda$ once in the counter-clockwise direction, as shown in Fig.~\ref{fig:cont_solns1}) 
\begin{equation}\label{eq:gen_closed_contour}
\Phi(\xi) = \oint dz~ e^{\xi z}(z-\lambda)^{\alpha_{+}-1}(z+\lambda)^{\alpha_{-}-1},
\end{equation}
which we evaluate by residues: \begin{equation}\label{eq:gen_res}
    \Phi(\xi) =\frac{2\pi i}{|\alpha_{-}|!} \left.\frac{d^{|\alpha_{-}|}}{dz^{|\alpha_{-}|}}\left[e^{\xi z}(z-\lambda)^{\alpha_{+}-1}\right]\right|_{z=-\lambda}.
\end{equation} 
From the Rodrigues formula for the associated (alternatively, "generalized") Laguerre polynomials~\cite{szego},
\begin{equation}\label{eq:rod_lag}
    L_N^{(b)}(\xi ) = \frac{1}{N!}e^\xi  \xi^{-b}\left .\frac{d^N}{dx ^N}\left(e^{-x }x^{N+b}\right)\right |_{x=\xi},
\end{equation}
we can see that the form of Eq.~(\ref{eq:gen_res}) is similar. To establish the exact correspondence, we define $u = -(z-\lambda)\xi $ (the point at which we evaluate the derivative, $z=-\lambda$, corresponds to $u = 2\lambda \xi $) and the inverse is $z=(\lambda\xi-u)/\xi$. Re-expressing the right-hand side as a function of $u$ yields \begin{eqnarray}
    \Phi(\xi) &= \nonumber \frac{2\pi i }{|\alpha_{-}|!}\left(-1\right)^{|\alpha_{-}|}\xi ^{|\alpha_{-}|}\left.\frac{d^{|\alpha_{-}|}}{du^{|\alpha_{-}|}}\left[e^{-u}e^{\lambda \xi }\left(-1\right)^{\alpha_{+}-1}\left(\frac{u}{\xi }\right)^{\alpha_{+}-1}\right]\right|_{u=2\lambda \xi }\\ \nonumber
    &=\frac{2\pi i (-1)^{\beta-1}
    }{|\alpha_{-}|!}\xi ^{-\beta+1}e^{\lambda \xi }\left.\frac{d^{|\alpha_{-}|}}{du^{|\alpha_{-}|}}\left[e^{-u}u^{\alpha_{+}-1}\right]\right|_{u=2\lambda \xi }\\ 
   &= \frac{2\pi i (-1)^{\beta-1}
    }{\left(2\lambda\right)^{-\beta+1}|\alpha_{-}|!}e^{-\lambda\xi }L_{|\alpha_{-}|}^{(\beta-1)}(2\lambda\xi).
\end{eqnarray}
Hence, except for a constant prefactor, which will be set via normalization, we find that 
\begin{equation}\label{eq:gen_laguerre_sol}
    \Phi(\xi) \propto e^{-\lambda \xi }L_{|\alpha_{-}|}^{(\beta-1)}\left(2\lambda \xi \right).
\end{equation}
Because the Laguerre polynomial of order $N$ goes like $\xi^N$ as  $\xi \to\pm\infty$, the asymptotic behavior of $\Phi(\xi)$ is dominated by the exponentially decaying term. Thus, this solution will be finite for all $0\leq \xi $ and is the wavefunction we sought to find. 

The last detail we need to work out is how the integer from the quantization condition $N=|\alpha_-|$, relates to the conventional principal quantum number $n$. In all harmonic oscillator cases, we have $n=N$, while the two dimensional Coulomb problem has $n-|m|-1=N$ and the three-dimensional Coulomb problem has $n-l-1=N$. These results are summarized, in terms of the principal quantum number, in Table~\ref{Tab:table3}.

\begin{table}[ht!]
\resizebox{\textwidth}{!}{
\centering
 \begin{tabular}{|c| c| c|c|} 
 \hline
 \multirow{2}*{\textbf{Problem}} &\textbf{Quantization}& \textbf{Energy}  & \multirow{2}*{\textbf{Form of $\Phi(\xi)$}} \\ &\textbf{Condition} & \textbf{Quantization, $E_n$}&\\  \hline
 
 1D SHO, & \multirow{2}*{$N=n = \frac{E}{2\hbar\omega}-\frac{1}{4}$} & \multirow{2}*{$\hbar\omega\left(2n+\frac{1}{2}\right) $} & \multirow{2}*{$e^{-\xi/2}L_n^{\left(-\frac{1}{2}\right)}(\xi)$}\\ Even &&&\\ \hline
 
 1D SHO, & \multirow{2}*{$N=n = \frac{E}{2\hbar\omega}-\frac{3}{4}$} & \multirow{2}*{$\hbar\omega\left(2n+1+\frac{1}{2}\right) $} & \multirow{2}*{$\xi^{1/2}e^{-\xi/2}L_n^{\left(\frac{1}{2}\right)}(\xi)$}\\ Odd &&&\\ \hline
 
 2D& \multirow{2}*{$N=n=\frac{E}{2\hbar\omega}-\frac{2|m|+1}{2}$} &\multirow{2}*{$\hbar\omega\left(2n+|m|+1\right)$} & \multirow{2}*{$e^{-\xi/2}L_{n}^{(|m|)}(\xi)$}\\SHO&&&\\ \hline
 
 3D  & \multirow{2}*{ $N=n=\frac{E}{2\hbar\omega}-\frac{1}{2}\frac{2l+3}{2}$} &  \multirow{2}*{$\hbar\omega\left(2n+l+\frac{3}{2}\right)$} & \multirow{2}*{$e^{-\xi/2} L_{n}^{\left (l+\frac{1}{2}\right )}(\xi)$} \\SHO&&&\\ \hline
 
 2D &\multirow{2}*{$N=n-|m|-1=\frac{\hbar}{a_0\sqrt{-2\mu E}}-\left(|m|+\frac{1}{2}\right)$}  &\multirow{2}*{$-\frac{\hbar^2}{2\mu a_0^2 \left(n-\frac{1}{2}\right)^2}$} & \multirow{2}*{$e^{-\xi}L_{n-|m|-1}^{(2|m|)}(\xi) $ }  \\ Coulomb &&&\\ \hline
 
 3D &\multirow{2}*{$N=n-l-1=\frac{\hbar}{a_0\sqrt{-2\mu E}}-\left(l+1\right)$}  & \multirow{2}*{$-\frac{\hbar^2}{2\mu a_0^2n^2}$} & \multirow{2}*{$e^{-\xi}L_{n-l-1}^{(2l+1)}(\xi) $ }  \\ Coulomb &&&\\ \hline
 
 Morse & \multirow{2}*{$N=n=\frac{\sqrt{2\mu V_0}-\sqrt{-2\mu E}}{a\hbar}$} & \multirow{2}*{$-\frac{a^2\hbar^2}{2\mu}\left(n-\frac{\sqrt{2\mu V_0}}{a\hbar}\right)^2$} & \multirow{2}*{$e^{-\xi/2}L_{n}^{(2n-2\delta-1)}(\xi) $} \\ Potential&&& \\
 \hline
 
 1D SHO &\multirow{2}*{$N=n = \frac{E}{\hbar\omega}-\frac{1}{2}$} & \multirow{2}*{$\hbar\omega\left(n+\frac{1}{2}\right) $} & \multirow{2}*{$H_n\left(\xi\right)$}\\ Method 2&&&\\ \hline
 
 \end{tabular}
 }

\caption{Quantization condition for the Laplace method (by tradition, the principal quantum number $n$ starts from 0 for all cases, except the Coulomb cases, where it starts from $|m|+1$ in two dimensions and from $l+1$ in three dimensions); $N$ is required to be a nonnegative integer from the quantization condition arising from Laplace's method.  By tracing back through the definitions of $\Phi$ and $\xi$ in each case, one obtains the standard wavefunctions for each problem, up to a normalization constant, that still needs to be determined. The $\delta$ in the index of the associated Laguerre polynomial in the last column of the Morse potential satisfies $\delta=\frac{\sqrt{2\mu V_0}}{a\hbar}$. All models, except for the Morse potential, have an infinite number of bound states. The Morse potential has a finite number, where we are required to have $n<\sqrt{2\mu V_0}/a \hbar$.}
\label{Tab:table3}

\end{table}

\section{Examples: Bound states for the simple harmonic oscillator}

Having now constructed the general methodology to solve bound-state problems using the Laplace method, we show the concrete details for how the method is used for the two-dimensional simple harmonic oscillator, which we treat in polar coordinates. The Schr\"odinger equation is solved first by separating variables, that is by letting $\psi(\rho,\phi) = R(\rho)\Phi(\phi)$. The solution for the angular function is 
\begin{equation}
    \Phi(\phi) = e^{im\phi},~m=\pm1,~\pm2,~...
\end{equation} 
as summarized in Table~\ref{Tab:table1}.
From this, we obtain a radial equation:
\begin{equation}
    R''(\rho) +\frac{1}{\rho}R'(\rho) + \left(\frac{2\mu E}{\hbar^2}-\frac{\mu^2\omega^2}{\hbar^2}\rho^2-\frac{m^2}{\rho^2}\right).
\end{equation}
Now, we define the new independent variable $\xi = \frac{\mu\omega}{\hbar}\rho^2$, and make the ansatz that $R(\xi) = \xi^{\frac{|m|}{2}}\Phi(\xi)$ (also summarized in Table~\ref{Tab:table1}). Making these substitutions we find the Laplace form of the radial equation as 
\begin{equation}
    \xi\Phi''(\xi)+(|m|+1)\Phi'(\xi)+\left(\frac{E}{2\hbar\omega}-\frac{1}{4}\xi\right)\Phi(\xi) = 0,
\end{equation}
as summarized in Table~\ref{Tab:table2}.
Hence,
\begin{equation}
\beta=|m|+1,~ 
\delta=\frac{E}{2\hbar\omega}, ~\text{and}~ \lambda = \frac{1}{2}.
\end{equation}
Next, we construct \begin{equation}
\label{eq:2dSHO_alphas}
\alpha_{+} = \frac{1}{2}\left(|m|+1+ \frac{E}{\hbar\omega}\right)~~\text{and}~~
\alpha_{-} = \frac{1}{2}\left(|m|+1- \frac{E}{\hbar\omega}\right).
\end{equation}
The quantization condition becomes 
\begin{equation}
    N = |\alpha_{-}|=n \Rightarrow E_n = \hbar\omega\left( 2n+|m|+1\right).
\end{equation}
From Eqs.~(\ref{eq:gen_laguerre_sol}) and (\ref{eq:2dSHO_alphas}), we find the desired solution to the differential equation is 
\begin{equation}
\Phi(\xi) = e^{-\frac{\xi}{2}}L_{n}^{\left(|m|\right)}\left(\xi\right)
\end{equation}
as summarized in Table~\ref{Tab:table3}.
This then yields the following for the full wavefunction:
\begin{equation}
    \psi_{n,m}(\rho,\phi) = \rho^{|m|}e^{-\frac{\mu\omega}{2\hbar}\rho^2}L_{n}^{\left(|m|\right)}\left(\frac{\mu\omega}{\hbar}\rho^2\right)e^{im\phi},
\end{equation}
up to a normalization constant, which has not yet been determined.

The solutions we derived for the one-dimensional simple harmonic oscillator may not look familiar to many. But they actually are the standard form, expressed in terms of a Gaussian multiplied by a Hermite polynomial, once we realize that there is an identity relating associated Laguerre polynomials to Hermite polynomials, given by  \begin{eqnarray}\label{eq:laguerre_to_hermite}
    H_{2n}(x) = \left(-4\right)^n n! L_n^{\left(-\frac{1}{2}\right)}\left(x^2\right) \\
    \nonumber H_{2n+1}(x) = 2\left(-4\right)^n n! x L_n^{\left(\frac{1}{2}\right)}\left(x^2\right),
\end{eqnarray} 
where $H_n$ is the physicist's form for the Hermite polynomial defined by the Rodrigues' formula 
\begin{equation}
    H_n(x) = \left(-1\right)^n e^{x^2}\frac{d^n}{dx^n}\left (e^{-x^2}\right ).
\end{equation} 

The remaining cases follow in a similar fashion and can be constructed by following the summarizing formulas in Tables~\ref{Tab:table1}--\ref{Tab:table3}.

What remains is for us to discuss the second way to solve the one-dimensional harmonic oscillator, which obtains the Hermite polynomials directly from their Rodrigues' formula after determining the residue at the pole. This method is summarized in the last row of Tables~\ref{Tab:table1}-\ref{Tab:table3} and it uses a different ansatz for the wavefunction, which also leads to a differential equation in the Laplace form, but a different one from all of the other cases. It is given by
 \begin{equation}\label{eq:hermite_eqn}
    \Phi''\left(\xi\right)-2\xi\Phi'\left(\xi\right)-2\alpha\Phi\left(\xi\right)=0,
\end{equation}
where $\alpha =\frac{1}{2}- \frac{E}{\hbar\omega}$; note we must have $E\ge 0$ (since the Hamiltonian is a positive semidefinite operator in this case), so $\alpha\le \frac{1}{2}$. In addition, we need to find a finite solution for all $-\infty< \xi< \infty$. We now go through the steps of the Laplace method. First, compute the polynomials 
\begin{eqnarray}
P(z) = z^2-2\alpha \\ Q(z) = -2z,
\end{eqnarray}
and then exponentiate the antiderivative of the ratio $P/Q$, which now includes only one power of $z$ raised to a potentially noninteger exponent. This
allows us to write the solution to Eq.~(\ref{eq:hermite_eqn}) as the contour integral \begin{equation}\label{eq:hermite_cont_int}
    \Phi(\xi) = \int_C dz~ e^{\xi z -\frac{z^2}{4}}z^{\alpha-1}. 
\end{equation}
Next, we restrict the exponent of $z$ such that $\alpha\notin\mathbb{Z}$. The branch points of the integrand in Eq.~(\ref{eq:hermite_cont_int}) are at $z = 0$ and at infinity. To construct a single branch of the integrand, we choose to draw the branch cut from $z=0\to z=-\infty$ along the negative real axis. 

\begin{figure}
    \centering
    \includegraphics[width=0.75\columnwidth]{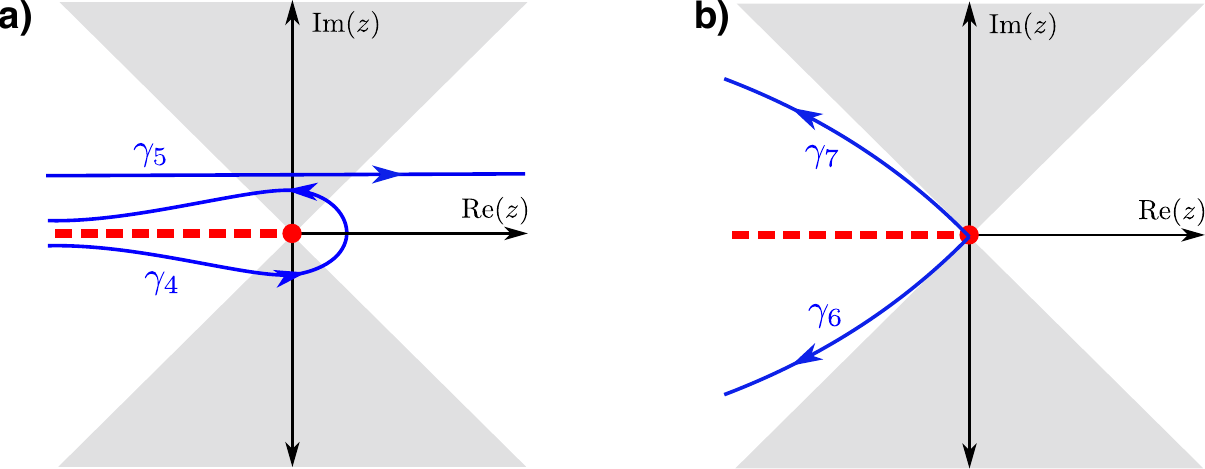}
    \caption{Four possible contours for solving the one-dimensional simple harmonic oscillator using the second ansatz. The contours must lie within the white ``cone'' regions as they go to infinity (there is no other restriction on the contours except not to cross the branch cut for finite values of $z$). In panel (a), we show the Hankel contour $\gamma_4$, which goes around the branch cut and $\gamma_5$, a contour that runs parallel to the real axis. In panel (b), we show two contours starting from the branch point and running to infinity either below ($\gamma_6$) or above ($\gamma_7$) the branch cut. }
    \label{fig:cont_1D_SHO}
\end{figure}

We next consider the possible contours as shown in Fig.~\ref{fig:cont_1D_SHO}. By the same arguments used before, we can eliminate all closed contours that don't contain a branch point or end at a branch cut and all open contours between any two finite points in the complex plane. Thus, the only candidate contours are ones with at least one endpoint at $z=0$ or that have endpoints that go to infinity. As the contour goes to infinity, it must remain inside a cone with an angle of $\pm\pi/4$ of the real axis, to be bounded. There is a cone around the positive real axis and also one around the negative real axis.

Thus, the contours to analyze are a Hankel-like contour around the branch cut, which we call $\gamma_4$, a contour from negative infinity to positive infinity (which does not cross the negative real axis), called $\gamma_5$, and a contour from the origin going to infinity with $\text{Re}(z)<0$, called $\gamma_6$, or from the origin to infinity with $\text{Re}(z)>0$, called $\gamma_7$; as long as we remain inside the cones about the real axis as we go to infinity. It is fairly easy to see any other contour with endpoints at $0$ and $\infty$ can be deformed into one of these contours or can be mapped to them by taking $z\to -z$. For example, the integral over $\gamma_7$ is converted into a contour from the branch point at 0 that runs to infinity inside the white cone along the positive real axis by transforming $z\to -z$, which is equivalent (up to a complex phase) to the integral over $\gamma_7$ but with $\xi\to-\xi$. 

The first condition we have is that the function $V(z)$ has identical values at the endpoints for all $\xi$. In this case, we have
\begin{equation}
    V(z) = e^{\xi z -\frac{z^2}{4}}z^{\alpha}.
\end{equation} 
It is easy to see that $\lim_{z\to 0}V(z)=\infty$ if $\alpha<0$, while $\lim_{z\to0}V(z)=0$ if $\alpha>0$. The asymptotic behavior of $V$ is dominated by the $e^{-z^2}$ term, when $|z|\to\infty$, so we have $\lim_{z\to\infty}V(z)=0$ when we lie inside the white cone about the real axes. This implies, that we can only have $z=0$ as an endpoint, when $\alpha>0$. But, in that case, any integral that has 0 as an endpoint diverges, due to the power-law behavior of the integrand near $z=0$ having too negative of an exponent. So no contour can have $z=0$ as an endpoint, eliminating $\gamma_6$ and $\gamma_7$.

The next condition to check the integral as $\xi\to -\infty$. Since the contour lies inside the cone, it seems like it will always be bounded, and should never diverge. But, in the region near $z\approx 2\xi$, the integrand actually behaves like $e^{\xi^2}$ and can give large contributions. The way to evaluate such situations is to perform an asymptotic analysis on the integrals to determine their value for large $|\xi|$. The standard way to do this is called the steepest-descents approach, which notes that the contributions to the exponential are largest near the maximum of the exponent, which is then described by a simple quadratic near the extremum, yielding a Gaussian integral that can be evaluated exactly. The analysis is straightforward, but is not often taught, so we describe it carefully, starting with the Hankel contour $\gamma_4$.  

The saddle-point approximation allows us to approximate integrals of the form
\begin{equation}
   \Phi(\xi) = \int_\gamma dz~e^{h(z)}g(z),
\end{equation}
which corresponds to the integral we need to evaluate for the Laplace method solution. Note that we have two options for how to proceed here. We can pick $g(z)=z^{\alpha-1}$, or we can pick $g(z)=1$ via writing $z^{\alpha-1}=e^{(\alpha-1)\ln z}$ and absorbing this term into the exponent $h(z)$. Both approaches yield the same final results, but the former is simpler than the latter, because there is only one saddle point in this case, making the analysis simpler. This means we have $h(z)=\xi z-\tfrac{1}{4}z^2$. Taking the derivative, to find the extrema, we have $h'(z)=\xi-\tfrac{1}{2}z$. Setting this equal to zero, tells us the extremum occurs at $z=z_0=2\xi$, which is called the saddle point; note that we have $h(z_0)=\xi^2$. In complex analysis, one direction through the saddle point is a minimum, while the other is a maximum, yielding a saddle-point shape for the exponent near the saddle point $z_0$. One must choose the direction for the contour through the saddle point to traverse the maximum, not the minimum. In this case, the maximum direction is along the real axis, which is simple to see, because we have a quadratic for the exponent with a negative real coefficient.

The asymptotic analysis next deforms the contour to go through the saddle point along the maximum direction. When $\xi<0$, this saddle point lies on the branch cut, so we will deform the Hankel contour to pass infinitesimally below it and parallel to the real axis---once below the negative real axis and once above. This yields two contributions for the steepest-descents integral. For the contribution from below the real axis, we parameterize the contour as given by $\gamma\approx z_0+t$ near the saddle point, so that
\begin{equation}
h(z)\approx h(z_0)-\frac{1}{4}t^2.
\end{equation}
Since the integrand decays quickly away from the saddle point, we extend the limits on $t$ to run from $-\infty$ to $\infty$ and we approximate $g(z)\approx g(z_0)$. This then gives us the contribution from the saddle point below the real axis to be
\begin{equation}
\int_{-\infty}^\infty dt e^{h(z_0)-\frac{1}{4}t^2}g(z_0)\approx 2\sqrt{\pi} (2|\xi|)^{\alpha-1}e^{-i\pi(\alpha-1)}e^{\xi^2}.
\end{equation}
The contribution from the saddle point above the negative real axis, is similar---it has an overall negative sign, because the contour runs from right to left instead of left to right and the sign of the phase in the exponent is positive because we are above the branch point. The total asymptotic estimate for the integral is then
\begin{equation}
    \int_{\gamma_4}dz e^{\xi z-\frac{1}{4}z^2}z^{\alpha-1}\approx -2^{\alpha+1}i\sqrt{\pi}|\xi|^{\alpha-1}\sin\pi(\alpha-1)e^{\xi^2}.
\end{equation}
Since we assume $\alpha$ is not an integer, the coefficient is nonzero and this gives a leading contribution that goes like $e^{\xi^2}$.
Looking at Table~\ref{Tab:table1}, we see that the full wavefunction is proportional to $e^{-\tfrac{1}{2}\xi^2}\Phi$, so this solution will go as $e^{\tfrac{1}{2}\xi^2}$ as $\xi\to-\infty$, which diverges. So, $\gamma_4$ does not yield a finite wavefunction. Moreover, a similar analysis yields the same asymptotic behavior for $\gamma_5$. Note that this analysis is similar to the Fr\"obenius analysis when the series does not truncate, and we reject the solution due to the wavefunction growing as we go to infinity.

So we do not obtain a finite solution from any of the possible contours, and we again must change our assumption that $\alpha\notin\mathbb{Z}$ to allow for a new contour. When $\alpha\in\mathbb{Z}$, the integrand is single-valued, so there no longer is a branch point or branch cut. Consequentially, we can now enclose the origin with a new closed contour. Our only choice is then that $\alpha\leq0,$ yielding the quantization condition given by \begin{equation}
    n = -\alpha=\frac{E}{\hbar\omega}-\frac{1}{2}\Rightarrow E_n = \hbar\omega\left(n+\frac{1}{2}\right),~n=0,1,2...
\end{equation}
This determines the energy levels of the one dimensional oscillator. Now, we write the (unnormalized) solution to the differential equation for this closed contour as
\begin{equation}\label{eq:oint_hermite}
    \Phi(\xi) = \oint dz~ e^{\xi z -\frac{z^2}{4}}z^{-n-1},~n=0,1,2\cdots,
\end{equation}
with a closed contour that encircles the origin.
By completing the square in the exponential term in Eq.~(\ref{eq:oint_hermite}), we can re-write this integral as 
\begin{equation}
    \Phi(\xi) = e^{\xi^2}\oint dz~ e^{-\left(\xi-\frac{z}{2}\right)^2}z^{-\left(n+1\right)}.
\end{equation}
Now, we let $u = \xi-\frac{z}{2}$, and by making this substitution, we obtain, up to a constant prefactor, 
\begin{equation}
    \Phi(\xi) = e^{\xi^2}\oint du ~ \left(-1\right)^n e^{-u^2} (u-\xi)^{-\left(n+1\right)}.
\end{equation}
This integral can be evaluated by residues about the pole of order $n+1$ at $u =\xi$:
\begin{equation}
     \Phi(\xi) = \left(-1\right)^n e^{\xi^2}\lim_{u\to\xi} \frac{d^n}{du^n}\left(e^{-u^2}\right) = \left(-1\right)^n e^{\xi^2} \frac{d^n}{d\xi^n}\left(e^{-\xi^2}\right).
\end{equation}
This is precisely the Rodrigues formula for the $n^\text{th}$ degree Hermite polynomial $H_n(\xi)$. That is, up to a constant prefactor, 
\begin{equation}
    \Phi(\xi) = H_n(\xi) = H_n\left(\sqrt{\frac{\mu\omega}{\hbar}}x \right), 
\end{equation}
which allows us to write the (unnormalized) wavefunction for the 1D harmonic oscillator as: 
\begin{equation}
    \psi(x) \propto e^{-\frac{\mu\omega}{2\hbar}x^2}H_n\left(\sqrt{\frac{\mu\omega}{\hbar}}x \right).
\end{equation}

\begin{table}[ht!]
\centering
 \begin{tabular}{|c| c| c|c|} 
 \hline
 \multirow{2}*{\textbf{Problem}} & \multirow{2}*{\textbf{Potential}}  & \textbf{Independent}&\textbf{Wavefunc.} \\&& \textbf{Variable}& \textbf{Form} \\ \hline
 
 2D Free& \multirow{2}*{$V =0$} & \multirow{2}*{$\xi = \sqrt{\frac{2\mu E}{\hbar^2}}\rho$}& \multirow{2}*{$\rho^{|m|}\Phi(\xi)e^{im\phi}$} \\Particle&&&\\ \hline
 
 3D Free& \multirow{2}*{$V = 0$} & \multirow{2}*{$\xi =\sqrt{\frac{2\mu E}{\hbar^2}}r$} & \multirow{2}*{$r^{l}\Phi(\xi)Y_l^m(\theta,\phi) $}\\ Particle &&&\\ \hline

 2D  &\multirow{2}*{$V=-\frac{e^2}{\rho}$}& \multirow{2}*{$\xi = \sqrt{\frac{2\mu E}{\hbar^2}}\rho$} & \multirow{2}*{$\rho^{|m|}\Phi(\xi)e^{im\phi}$} \\ Coulomb &&&\\ \hline
 
 3D  &\multirow{2}*{$V=-\frac{e^2}{r}$}& \multirow{2}*{$\xi = \sqrt{\frac{2\mu E}{\hbar^2}}r$}& \multirow{2}*{$r^l\Phi(\xi)Y_l^m(\theta,\phi)$}  \\ Coulomb &&&\\ \hline
 
 Morse  & \multirow{2}*{$V= V_0\left(e^{-2a x}-2e^{-a x}\right)$} & \multirow{2}*{$\xi =  \frac{2\sqrt{2\mu V_0}}{a\hbar}e^{-a x}$} & \multirow{2}*{$ \xi^{i\frac{\sqrt{2\mu E}}{a\hbar}}\Phi(\xi)$} \\ Potential&&& \\
 \hline
 
 \end{tabular}
\caption{Summary for how to convert the Schr\"odinger equation into the Laplace equation for the five problems with continuum solutions. For each, we give the form of the potential, the general form of the wavefunction where $\Phi$ is the part of the solution that is found by using the Laplace method, and the independent variable $\xi$ used for each problem. Note that $E>0$ for these continuum problems. We have $m$ denoting the quantum number for the $z$-component of angular momentum and  $l$ denoting the quantum number for the total angular momentum.} 
\label{Tab:table4}
\end{table}

\section{Continuum solutions with the Laplace method}
There are several quantum systems whose energy eigenstates have energy eigenvalues that lie in the continuum and that we can also treat with the Laplace method; this includes the free particle in two, and three dimensions, the continuum solutions of the Coulomb problem in two- and three-dimensions, and the continuum solutions of the one-dimensional Morse potential. The steps for obtaining a differential equation in the Laplace form are similar to what we already showed above, and in Table~\ref{Tab:table4} we summarize the results for these different models. Note that the substitutions required for the free-particle problems look like the final wavefunction will diverge at the origin, but we require the function $\Phi(\xi)$ to have a high-enough order zero at the origin that the final wavefunction remains finite everywhere. In Table~\ref{Tab:table5}, we show the final differential equations obtained by this procedure, which is similar to the bound-state form in Eq.~(\ref{eq:laguerre_diffeq_form}), but with the sign of the $\bar{\lambda}^2$ term changed. We have written that term as $-(i\bar{\lambda})^2$ instead of $+\bar{\lambda}^2$ to simplify the notation that we need in solving the problem. The Morse potential, on the other hand, keeps the form of Eq.~(\ref{eq:laguerre_diffeq_form}), but some parameters now become complex.

\begin{table}[ht!]
\centering

 \begin{tabular}{|c| c| c|} 
 \hline
 \multirow{2}*{\textbf{Problem}} & \textbf{Laplace Form of the} &\multirow{2}*{$\alpha_{\pm}$} \\ & \textbf{Schr\"odinger Equation}& \\ \hline

 2D Free & \multirow{2}*{$\xi\Phi''+(2|m|+1)\Phi'+\xi\Phi=0$}&\multirow{2}*{$|m|+\frac{1}{2}$} \\ Particle &&\\
 \hline
 
 3D Free & \multirow{2}*{$\xi\Phi''+2(l+1)\Phi'+\xi\Phi=0$}& \multirow{2}*{$l+1$} \\ Particle&&\\
 \hline
 
 2D   & \multirow{2}*{$\xi\Phi''+(2|m|+1)\Phi'+\left(\frac{2\hbar}{a_0\sqrt{2\mu E}} + \xi\right)\Phi=0$} & \multirow{2}*{$|m|+\frac{1}{2}\mp\frac{i\hbar}{a_0\sqrt{2\mu E}}$}\\Coulomb&&\\
 \hline
 
 3D   & \multirow{2}*{$\xi\Phi''+2(l+1)\Phi'+\left(\frac{2\hbar}{a_0\sqrt{2\mu E}} + \xi\right)\Phi=0$}& \multirow{2}*{$l+1\mp\frac{i\hbar}{a_0\sqrt{2\mu E}}$} \\Coulomb&&\\
 \hline
 
 Morse & \multirow{2}*{$\xi\Phi''+\left(2i\frac{ \sqrt{2\mu E}}{a\hbar}+1\right)\Phi' + \left(\frac{\sqrt{2\mu V_0}}{a\hbar}-\frac{1}{4}\xi\right)\Phi=0$}&\multirow{2}*{$\frac{ i\sqrt{2\mu E}\pm\sqrt{2\mu V_0}}{a\hbar}+\frac{1}{2}$} \\ Potential&& \\
 \hline
 
\end{tabular}
\caption{Final differential equation and exponents $\alpha_{\pm}$ for continuum cases to be solved by the Laplace method. The second column summarizes the final form of the Schr\"odinger equation obtained by the substitutions detailed in Table~\ref{Tab:table4}. Note that because $\delta=0$ for the free-particle cases, there is only one exponent for those problems. In all problems except the Morse potential, we have $\bar{\lambda}=1$.}  
\label{Tab:table5}
\end{table}

We begin with the free particle and Coulomb problems, each treated in both two and three dimensions, because they all are treated similarly. The differential equation in the Laplace form takes the form 
\begin{equation}
\label{eq:laguerre_diffeq_form2}
\xi\Phi''(\xi) + \beta \Phi'(\xi) +\left(\delta-(i\bar{\lambda})^2\xi\right)\Phi(\xi) = 0,
\end{equation}
for all of these cases.
This means that we can write the solution as 
\begin{equation}\label{eq:continous_soln_int}
    \Phi(\xi) = \int_\gamma dz~e^{\xi z}\left(z-i\bar{\lambda}\right)^{\alpha_+-1}\left(z+i\bar{\lambda}\right)^{\alpha_- -1},
\end{equation}
by following the Laplace method. Here, we have
\begin{equation}
    \alpha_{\pm} = \frac{i\beta\bar{\lambda}\pm \delta}{2i\bar{\lambda}}.
\end{equation}
In each of these four cases, $\bar{\lambda}=1$. For the free-particle problems, we have $\delta=0$, so there is only one $\alpha$.

\begin{figure}
    \centering
    \includegraphics[width=0.45\columnwidth]{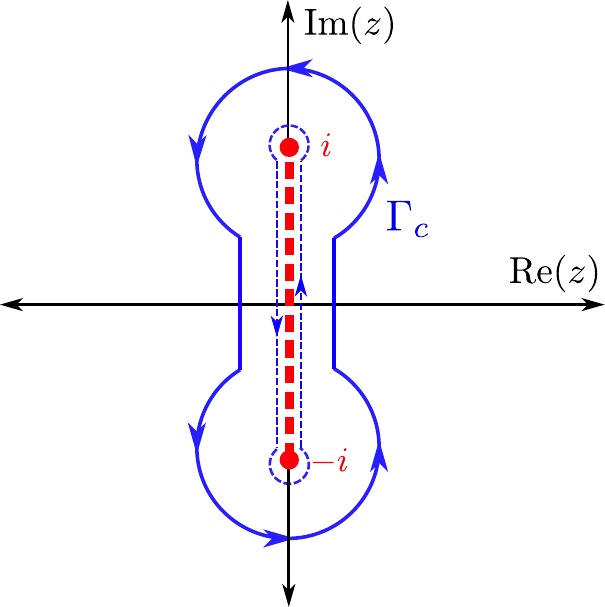}
    \caption{Rotated 'dog-bone' shaped contour for evaluating the contour integral for the continuum wave functions of the Coulomb problem in two and three dimensions.}
    \label{fig:cont_solns2}
\end{figure}

As with bound states, we restrict the contours over which we integrate by requiring that the wavefunction be finite everywhere. As discussed in \cite{our_paper}, this constraint yields a rotated dog-bone shaped contour which encloses the branch points $\pm i$, as seen in Fig.~\ref{fig:cont_solns2}. The reason why is that any contour that runs out to infinity will yield an infinite result for $\Phi(0)$.  We do, however, mention that the case of a 3d free particle actually does not need the branch cut, because both $\alpha_\pm$ are positive integers, and the integrand is not multivalued. In this case, the integral must be taken along the imaginary axis from $-i$ to $i$, connecting the two points where $V=0$ for all $\xi$. Note that this choice could also have been used for the Coulomb potential cases (because $V=0$ at the branch points for all $\xi$ as well), but there is pedagogical value to using the dog-bone contour because it allows us some additional options for evaluating $\Phi(\xi)$ numerically, as we discuss below; since the solutions corresponding to both choices of the contour are proportional to each other (shown below), we can freely choose either one. The ambiguity in the prefactor is always removed when the final wavefunction is normalized (but we will not discuss normalization in this work).

\begin{figure}
    \centering
    \includegraphics[width=\columnwidth]{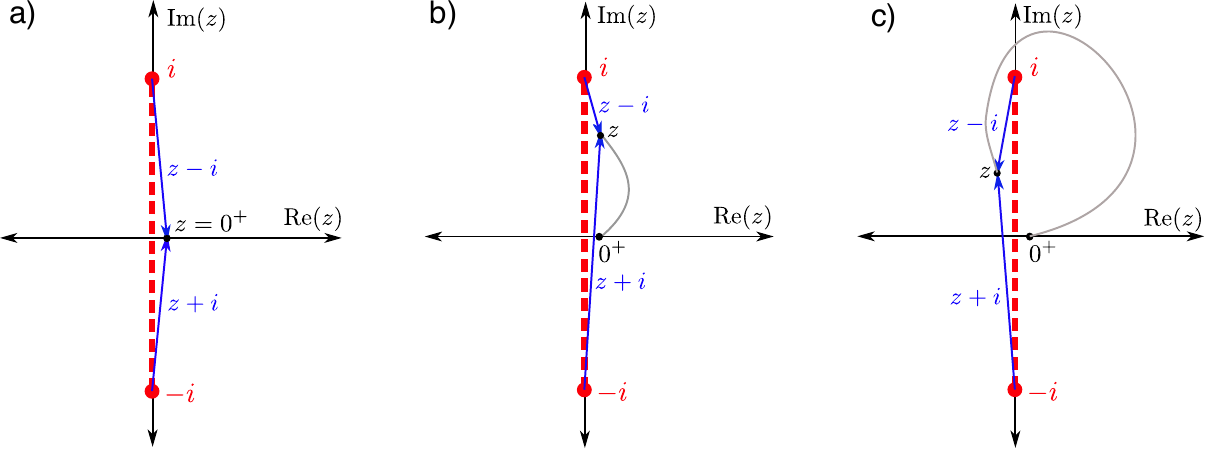}
     \caption{Procedure for determining the phases of the integrand in Eq.~(\ref{eq:continous_soln_int}) along the vertical pieces of the contour (we deformed the contour slightly for clarity in the image). a) We pick a reference point $z=0^+$ and show arrows from $i$ and $-i$ to the reference point. b) We draw paths from the reference point $0^+$ to each point $z$ of this piece of the deformed contour $\Gamma_c$ (here running upward vertically to the right of the branch cut). The arrows drawn from the upper and lower branch points to $z$ do not change their net direction as they move along the path from $0^+$ to $z$, thus the phase for $f(z)$ is the same as the phase for $f(0^+)$ on the right side of the branch cut. c) For reaching the points of the contour on the left side of the branch cut, the arrow drawn from $i$ to $z$ needs to rotate by $2\pi$, while the net change in direction of the arrow drawn from the lower branch point is zero; note that the arrow along the path from the reference point to $z$ is allowed to cross the branch cut, even though the path never crosses the branch cut.}
    \label{fig:phases}
\end{figure}

It is critical to evaluate the phases properly, when determining the integrand. With the branch cut structure we use, the function is single-valued everywhere in the complex plane, except along the branch cut itself, but the function has different values on both sides of the branch cut. We start by picking a reference point, which will be the origin, just to the right of the branch cut, which we call $z=0^+$, as shown in panel (a) of Fig.~\ref{fig:phases}. The multivalued function in the integrand is $f(z)=(z-i)^{\alpha_+-1}(z+i)^{\alpha_--1}$ and we focus on how to determine the phase consistently for this function. We find that $f(0^+)=(-i)^{\alpha_+-1}(i)^{\alpha_--1}$, which we evaluate with the standard phases $i=e^{i\frac{\pi}{2}}$ and $(-i)=e^{-i\frac{\pi}{2}}$. Then we have $f(0^+)=\exp\Big (i\frac{\pi}{2}(\alpha_--\alpha_+)\Big )$. Noting the form of $\alpha_\pm$, we find $f(0^+)=\exp(-\frac{\pi}{2}\delta)$.

Now, to calculate $f(z)$ anywhere in the complex plane, we first draw a path from  the reference point $0^+$ to $z$ that does not cross the branch cut (see panels (b) and (c) of Fig.~\ref{fig:phases}). Then, we examine how arrows drawn from the upper branch point $i$ to a point on the path rotate as we move along the path from $0^+$ to $z$. This determines the change in the phase for the factor $(z-i)$. We repeat with an arrow drawn from $-i$ to a point along the path, and follow it from $0^+$ to $z$. The rotation of the arrow here, again determines the change in the phase for the factor $(z+i)$. Each of those factors will have the change in phase multiplied by the corresponding exponent, and that will determine the phase of $f(z)$. We will show below that using this procedure produces a single-valued function over the entire complex plane. But first, we will use this procedure to convert the integral form of our solution to a single integral that runs along the real axis.

We deform the contour in Fig.~\ref{fig:cont_solns2} to be infinitesimally close to the imaginary axis along the vertical portions, and wrapping infinitesimally close to the branch points in the circular portions, we obtain a result expressed as the sum of the contributions from two vertical lines (downward from $i\to -i$ just to the left of the axis and upward from $-i\to i$ to the right of the axis) and from the infinitesimal circular arcs enclosing the two branch points. It is easy to see that the contributions from the circular arcs around the branch points will be zero. Since $\text{Re}(\alpha_{\pm})>0$, the integral around each arc will go to zero, so the branch points do not contribute to the integral. The integral over the full contour then reduces to the sum of the contributions from the two vertical lines. Note that since they lie on either side of the branch cut, there is a phase difference between the two integrals, and the contributions will not cancel. 

To determine this phase difference, we follow the procedure described above (see Fig.~\ref{fig:phases} for a graphical representation). We start with the piece of the contour that runs from $-i$ to $i$ along the right hand side of the imaginary axis. We can draw a path from the reference point at $0^+$ to any point along this piece of the deformed contour as a straight vertical line. We can immediately see that the arrow from $i$ to $z$ on the contour does not change direction after moving along the path. Neither does the arrow drawn from $-i$ to $z$. This means the phase for $f(z)$ is the same as the phase for $f(0^+)$ along the entire piece of the contour. We write $z=iy$, with $y$ real, and this part of the contour integral becomes the following:
\begin{equation}
I_1=\int_{-1}^1 i\,dy\, e^{iy\xi}|y-1|^{\alpha_+-1}|y+1|^{\alpha_--1}e^{-\frac{\pi}{2}\delta}.
\end{equation}
The second piece of the contour runs along the left hand side of the imaginary axis from $i$ to $-i$. We draw a path from the reference point $0^+$ to $z$ along the contour, by an arc that goes around the upper branch point at $i$. One can immediately see that an arrow drawn from $i$ to the reference point $0^+$, winds by $2\pi$ in the counterclockwise direction as it goes around the path to $z$. So the change in the phase for the factor $(z-i)$ is $2\pi$. The arrow drawn from $-i$ to the reference point, will rotate first to the right, and then to the left, but it ultimately accumulates no net phase, so its change in phase is $0$. The fact that the final point ended on the left-hand side of the branch point does not determine the change in the phase along the path, it is the net motion of the arrow as we follow along the path that does this. This point can often be misunderstood by students. It is important to note that our rule for determining the phase of the function does not input by hand a change of phase when crossing the branch cut. Instead, we follow the described procedure to determine the change in the phases. 
Again, we let $z=iy$ and we note that the integral runs down the imaginary axis, so we find the contribution from this piece of the contour is the following:
\begin{equation}
    I_2=\int_1^{-1} i\,dy\,e^{i\xi y}|y-1|^{\alpha_+-1}|y+1|^{\alpha_--1}e^{2i\pi(\alpha_+-1)}e^{-\frac{\pi}{2}\delta}.
\end{equation}
The factor $e^{-2i\pi}=1$ and can be ignored. The real part of $\alpha_+$ is either an integer or a half-odd integer. In the former case, the added factor is $1$ and can be ignored, in the latter case it is $-1$. The imaginary part of $\alpha_+$ adds in a factor of $e^{\pi\delta}$. Hence, when we combine the two integrals together (and recall that we have to switch the order of the limits in the second integral) we find the total contour integral becomes
\begin{equation}
    \Phi(\xi)=i\left (e^{-\frac{\pi}{2}\delta}\mp e^{\frac{\pi}{2}\delta}\right )\int_{-1}^1 dy\, e^{i\xi y}|1-y|^{\alpha_+-1}|1+y|^{\alpha_--1},
\end{equation}
where the minus sign is for when the real part of $\alpha_+$ is an integer and the plus sign is for when it is a half-odd integer.

To convert this into a form that is easily expressed in terms of confluent hypergeometric functions, we let $y=-1+2x$ and substitute into the integral to find that
\begin{equation}
    \label{eq:real_int}
    \Phi(\xi)=i\left (e^{-\frac{\pi}{2}\delta}\mp e^{\frac{\pi}{2}\delta}\right )2^{\alpha_++\alpha_--1}e^{-i\xi}\int_{0}^1 dx\, e^{2i\xi x}|1-x|^{\alpha_+-1}x^{\alpha_--1}.
\end{equation}
Comparing to the standard integral form of the Kummer function (for $\text{Re}({b})>\text{Re}({a})>0$) (as given by Eq.~13.4.1 of Ref.~\cite{dlmf})
\begin{equation}
    M(a,b,z)=\frac{\Gamma(b)}{\Gamma(a)\Gamma(b-a)}\int_0^1dx\, e^{zx}x^{a-1}(1-x)^{b-a-1},
    \label{eq:kummer_int_form}
\end{equation}
we find that
\begin{equation}
    \Phi(\xi)=i\left (e^{-\frac{\pi}{2}\delta}\mp e^{\frac{\pi}{2}\delta}\right )2^{\alpha_++\alpha_--1}\frac{\Gamma(\alpha_+)\Gamma(\alpha_-)}{\Gamma(\alpha_++\alpha_-)}e^{-i\xi}M(\alpha_-,\alpha_++\alpha_-,2i\xi).
    \label{eq:phi_kummer}
\end{equation}
Note that for $E>0$, the numerical prefactor is never zero, so we can always remove it from further discussion in the summary of the wavefunctions. It will enter, and is important, when we evaluate the results numerically below. One does need to complete the normalization step for the final wavefunctions (which is usually done with delta-function normalization), but we will not discuss that further here and instead will only summarize unnormalized wavefunctions, with the prefactor removed. Note that the result here corrects a sign error in the final result for the continuum wavefunction in Ref.~\cite{our_paper} arising from an inconsistent definition of the phase of the multivalued function.

We comment briefly here on the 3d free particle. This case results in just the $I_1$ term, which has the same final form as we have for the cases with the ``dog-bone'' contour (just with a different prefactor). So its result falls into the same category as the other three cases. It is just that we do not need to worry about any phase issues in working with the integrand in this case. Because there is no branch point, we cannot describe that solution with a closed contour, because such an integral always vanishes. Instead, we simply integrate from one ``zero'' point to the other.

One other point to note, is that using the Kummer relation (Eq.~13.2.39 of Ref.~~\cite{dlmf})
\begin{equation}
    M(a,b,z)=e^zM(b-a,b,-z),
\end{equation}
we can show that the unnormalized $\Phi(\xi)$, given by $e^{-i\xi}M(\alpha_-,\alpha_++\alpha_-,2i\xi)$, is real for real $\xi$. In particular, we have
\begin{equation}
    \left ( e^{-i\xi}M(\alpha_-,\alpha_++\alpha_-,2i\xi)\right )^*=e^{i\xi}M(\alpha_-^*,\alpha_++\alpha_-,-2i\xi),
\end{equation}
where we used the facts that $\alpha_++\alpha_-$ and $\xi$ are both real. Then, if you note that $\alpha_-^*=\alpha_+=(\alpha_++\alpha_-)-\alpha_-$, and use the Kummer relation, the right hand side of the equation becomes $e^{i\xi}e^{-2i\xi}M(\alpha_-,\alpha_++\alpha_-,2i\xi)$, which is equal to the original function and hence shows that this combination is real. Thus, if we drop the constant prefactor, the unnormalized $\Phi(\xi)$ can always be chosen to be real-valued. Note as well that at $\xi=0$, we have $\Phi(\xi)=1$, because $M(a,b,0)=1$ for all cases where the Kummer function is well defined from its power-series, which is the situation we have here. So this choice of contour leads to the correct continuum wavefunction, as we claimed earlier.
A summary of the unnormalized wavefunctions for all potentials (which have \textit{real} radial functions) is given in Table~\ref{Tab:table6}. These results follow by simply plugging in the explicit values of $\alpha_\pm$ and noting that there are identities between confluent hypergeometric functions and other functions, such as Bessel functions and spherical Bessel functions (as summarized in section 13.6 of the NIST Digital Library of Mathematical Functions~\cite{dlmf}). We do not show the details for how to carry out that algebra here.

One of the benefits of using contour-integral representations for the continuum wavefunctions is that it allows us to explore different ways to determine the wavefunctions numerically. For example, in this case we have three equivalent numerical representations. The first involves a real integral and is given in Eq.~(\ref{eq:phi_kummer}). The second involves our original integral representation in Eq.~(\ref{eq:laguerre_diffeq_form2}), where, for concreteness, we use a circular contour of radius $R$ centered at the origin for evaluating the wavefunction. The third is to develop a power series representation and then to numerically evaluate the series. This is done by deforming the contour until it has a very large radius, and then extracting the residue at infinity.

We describe how to determine this power series next. Key to completing the calculation is the determination of the Laurent series near the point at infinity. This is most easily determined by approaching infinity along the positive imaginary axis; we add a constant shift by $-i$ first, because we know the final result has a factor of $e^{-i\xi}$. So, we let $z=-i+i/y$ for $y$ a positive real number near zero. The phase for the factor $z-i$ is $i\pi$, because the arrow from the reference point wraps by $\pi$ as we move up the imaginary axis, while the phase from $z+i$ is zero. This gives us an overall phase factor of $-e^{i\pi\alpha_+}$. The integrand (including the change of variables factor $-i/y^2$) is then given by
\begin{equation}
    ie^{i\pi\alpha_+-\frac{\pi}{2}\delta}e^{-i\xi}e^{i\frac{\xi}{y}}\frac{1}{y^{\alpha_++\alpha_-}}|1-2y|^{\alpha_+-1}
\end{equation}
(the term $e^{-\frac{\pi}{2}\delta}$ comes from the phase of the function at our reference point).
We now want to expand this in a Laurent series for small $y$. Using the generalized binomial theorem for complex powers, yields
\begin{equation}
    ie^{i\pi\alpha_+-\frac{\pi}{2}\delta}e^{-i\xi}\sum_{m=0}^\infty \frac{(i\xi)^m}{m!}\sum_{j=0}^\infty \frac{(\alpha_+-1)_j}{j!}(-2)^jy^{j-m-\alpha_+-\alpha_-},
    \label{eq:laurent}
\end{equation}
where $(\alpha-1)_j$ is the Pochammer symbol for the falling factorial, given by $(\alpha-1)(\alpha-2)\cdots(\alpha-j)$. If we now perform a contour integral around the point at infinity, the result will be given by the residue, which is determined by the coefficient of the expansion in Eq.~(\ref{eq:laurent}) of the term $1/y$. Since $\alpha_++\alpha_-$ is an integer, we can immediately determine the residue. It is given by
\begin{equation}
    \text{Residue}=ie^{i\pi\alpha_+-\frac{\pi}{2}\delta}e^{-i\xi}\sum_{j=\alpha_++\alpha_--1}^\infty \frac{(\alpha_+-1)_j}{j!}(-2)^j\frac{(i\xi)^{j+1-\alpha_+-\alpha_-}}{(j+1-\alpha_+-\alpha_-)!}.
\end{equation}
Shifting the summation to start from zero, then gives
\begin{equation}
    \text{Residue}=ie^{i\pi\alpha_+-\frac{\pi}{2}\delta}e^{-i\xi}(-2)^{\alpha_++\alpha_--1}\sum_{j=0}^\infty \frac{(\alpha_+-1)_{j+\alpha_++\alpha_--1}}{(j+\alpha_++\alpha_--1)!}\frac{(-2i\xi)^{j}}{j!}.
\end{equation}
The confluent hypergeometric functions are typically expressed in terms of the rising factorials, instead of the falling factorials. Converting between them gives 
\begin{equation}
    (\alpha_+-1)_{j+\alpha_++\alpha_--1}=(-1)^{j+\alpha_++\alpha_--1}(1-\alpha_+)^{(j+\alpha_++\alpha_--1)}.
\end{equation}
This can then be expressed in terms of Gamma functions as 
\begin{equation}
    (\alpha_+-1)_{j+\alpha_++\alpha_--1}=(-1)^{j+\alpha_++\alpha_--1}\frac{\Gamma(j+\alpha_-)}{\Gamma(1-\alpha_+)}=(-1)^{j+\alpha_++\alpha_--1}(\alpha_-)^{(j)}\frac{\Gamma(\alpha_-)}{\Gamma(1-\alpha_+)}.
\end{equation}
The term in the denominator can be written as
\begin{equation}
    (j+\alpha_++\alpha_--1)!=(\alpha_++\alpha_-)^{(j)}\Gamma(\alpha_++\alpha_-).
\end{equation}
This means we have established that
\begin{equation}
    \frac{(\alpha_+-1)_{j+\alpha_++\alpha_--1}}{(j+\alpha_++\alpha_--1)!}=(-1)^{j+\alpha_++\alpha_--1}\frac{(\alpha_-)^{(j)}}{(\alpha_++\alpha_-)^{(j)}}\frac{\Gamma(\alpha_-)}{\Gamma(1-\alpha_+)\Gamma(\alpha_++\alpha_-)}.
\end{equation}
Using Euler's reflection formula $\Gamma(z)\Gamma(1-z)=\pi/\sin(\pi z)$ gives our final result:
\begin{equation}
    \text{Residue}=e^{-\frac{\pi}{2}\delta}\frac{e^{2i\pi\alpha_+}-1}{4\pi}2^{\alpha_++\alpha_-}\frac{\Gamma(\alpha_+)\Gamma(\alpha_-)}{\Gamma(\alpha_++\alpha_-)}e^{-i\xi}\sum_{j=0}^\infty\frac{(\alpha_-)^{(j)}}{(\alpha_++\alpha_-)^{(j)}}\frac{(2i\xi)^j}{j!}.
\end{equation}
Multiplying by $-2\pi i$ to determine the integral via the calculus of residues then produces a result equal to that in Eq.~(\ref{eq:phi_kummer}), provided
\begin{equation}
    M(\alpha_-,\alpha_++\alpha_-,2i\xi)=\sum_{j=0}^\infty\frac{(\alpha_-)^{(j)}}{(\alpha_++\alpha_-)^{(j)}}\frac{(2i\xi)^j}{j!},
\end{equation}
which is the standard definition of the confluent hypergeometric function, as long as $\alpha_++\alpha_-$ is not a nonpositive integer. This provides a numerical way to evaluate the wavefunction via a power series expansion. 

We can also evaluate the wavefunction using a simple numerical integration over the two different integral formulas for $\Phi$. The real-valued integral in Eq.~(\ref{eq:real_int})  is straightforward to evaluate, with the caveat that one evaluates the complex exponentials carefully, noting that $x^{a+ib}=x^ae^{ib\ln x}$, for example. Care must also be taken when the real part of the exponents are less than one, because those functions are not so easily integrated using traditional integration rules, without changing variables to remove their nonpolynomic behavior first.

\section{Contour integral around circular path for the Coulomb problem}

\begin{figure}
    \centering
    \includegraphics[width=0.8\columnwidth]{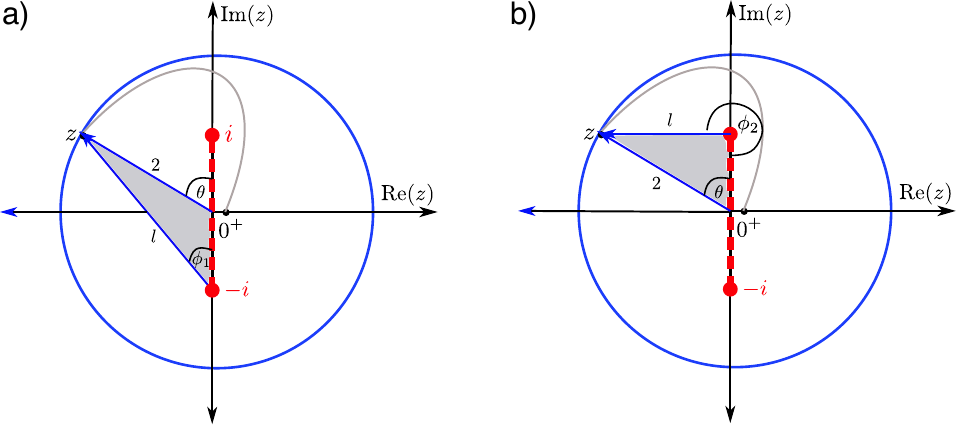}
     \caption{Geometry for determining the relationship between the phases $\phi_1$ and $\phi_2$ and $\theta$. In the figures, the symbol $l$ is used for the unknown length on each triangle and $R$ is chosen to equal 2 for concreteness. This case corresponds to $0\le\theta\le\pi$.}
    \label{fig:circ_phi1}
\end{figure}

For the contour integral around the circular path (see Eq.~(\ref{eq:laguerre_diffeq_form2})), we need to carefully determine the phase of the multivalued function in the integrand again. We use our standard approach, relating everything to our reference point.
With the angle $\theta$ measured relative to the positive imaginary axis, the parametrization of the radius $R$ integral is $z=Re^{i(\theta+\frac{\pi}{2})}=R(-\sin\theta +i\cos\theta)$; for concreteness, we will pick $R=2$ in the figures. Note that the conventional angle for describing $z$ in the complex plane is measured from the real axis, hence, the polar angle for $z$ is $\theta+\tfrac{\pi}{2}$. The integral then becomes
\begin{align}
    \Phi(\xi) &= i\int_0^{2\pi}d\theta\,Re^{i(\theta+\tfrac{\pi}{2})} e^{R\xi(-\sin\theta+i\cos\theta)}\left |Re^{i(\theta+\frac{\pi}{2})}-i\right |^{\alpha_+-1} \left |Re^{i(\theta+\frac{\pi}{2})}+i\right |^{\alpha_--1} \nonumber\\
    &~~~~~~~~~~\times e^{i\phi_2(\alpha_+-1)}e^{i\phi_1(\alpha_--1)}e^{-\frac{\pi}{2}\delta}.
\end{align}
Now we need to determine the phases $\phi_1$ and $\phi_2$ as a function of $\theta$. The graphics in Fig.~\ref{fig:circ_phi1} are helpful for this task, and represents the situation when $0\leq\theta < \pi$.  
The phase $\phi_1$ corresponds to the angle, which an arrow drawn from $-i$ to $0^+$ winds when it moves along the indicated path from the reference point $0^+$ to a  point on the circle (Fig.~\ref{fig:circ_phi1}a). From the angles and sides of the shaded triangle, we can extract the relation between $\phi_1$ and $\theta$. We employ the law of cosines
\begin{equation}
    l^2= R^2+1-2R\cos(\pi-\theta)=R^2+1+2R\cos\theta,
\end{equation}
and the law of sines
\begin{equation}
    \frac{\sin\phi_1}{R}=\frac{\sin(\pi-\theta)}{l}=\frac{\sin\theta}{l},
\end{equation}
and thus find
\begin{equation}
\label{eq:phi1}
    \sin\phi_1=\frac{R\sin\theta}{\sqrt{R^2+1+2R\cos\theta}}.
\end{equation}
For extracting $\phi_1$, we need to take the $\arcsin$ of the right hand side, and be careful from which quadrant to choose $\phi_1$ (will be discussed in detail later). Note that the final result looks like we just drew the angle ignoring the branch cut, but we did carefully follow the procedure of traversing a path that does not cross the branch cut, as required for the determination of the angle.

The phase $\phi_2$ we determine in a very similar way. This phase $\phi_2$ is the winding angle of an arrow drawn from $+i$ to $0^+$, which winds along the indicated path from the reference point $0^+$ to a point on the circle (see Fig.~\ref{fig:circ_phi1}b). Note that for $\theta=0$, the arrow has already flipped, i.e. rotated by $\pi$. The corresponding angle in the shaded triangle is thus $2\pi-\phi_2$. Again, we employ the law of cosines
\begin{equation}
    l^2=R^2+1-2R\cos\theta=R^2+1-2R\cos\theta,
\end{equation}
and the law of sines
\begin{equation}
     \frac{\sin\theta}{l} = \frac{\sin(2\pi-\phi_2)}{R} = \frac{-\sin\phi_2}{R} 
\end{equation}
Thus, we find
\begin{equation}
\label{eq:phi2}
    \sin\phi_2=\frac{-R\sin\theta}{\sqrt{R^2+1-2R\cos\theta}}.
\end{equation}

\begin{figure}
    \centering
    \includegraphics[width=0.8\columnwidth]{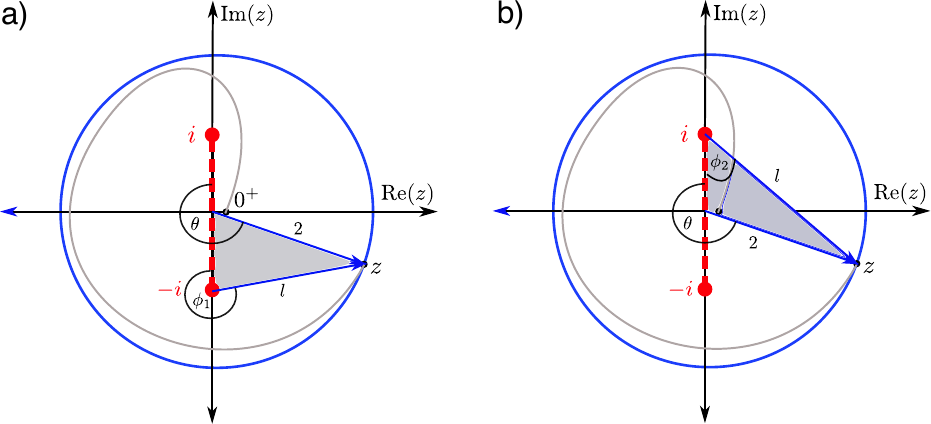}
     \caption{Similar figure as in Fig.~\ref{fig:circ_phi1}, except here we have $\pi\le\theta\le 2\pi$.}
    \label{fig:circ_phi2}
\end{figure}
Now we repeat the same procedure for $\pi\leq\theta <2\pi$. The corresponding plots for determining $\phi_1$ and $\phi_2$ are shown in Fig.~\ref{fig:circ_phi2}. In Fig.~\ref{fig:circ_phi2}a, the angles in the shaded triangle are $\theta-\pi$ and $2\pi-\phi_1$. The law of cosines 
\begin{equation}
    l^2=R^2+1-2R\cos(\theta-\pi)= R^2+1+2R\cos\theta,
\end{equation}
and the law of sines
\begin{equation}
    \frac{\sin(2\pi-\phi_1)}{R}=\frac{-\sin\phi_1}{R}=\frac{\sin(\theta-\pi)}{l}=\frac{-\sin\theta}{l},
\end{equation}
yield
\begin{equation}
    \sin\phi_1=\frac{R\sin\theta}{\sqrt{R^2+1+2R\cos\theta}},
\end{equation}
which is the same as Eq.~(\ref{eq:phi1}).
For $\phi_2$, the geometric relation are shown in Fig.~\ref{fig:circ_phi2}b. Again, we use the law of cosines
\begin{equation}
    l^2=R^2+1-2R\cos(2\pi-\theta)=R^2+1-2R\cos\theta,
\end{equation}
and the law of sines
\begin{equation}
    \frac{\sin(2\pi-\theta)}{l}=\frac{-\sin\theta}{l}=\frac{\sin\phi_2}{R}
\end{equation}
and obtain
\begin{equation}
    \sin\phi_2=\frac{-R\sin\theta}{\sqrt{R^2+1-2R\cos\theta}},
\end{equation}
which is the same as Eq.~(\ref{eq:phi2}). Thus, the same formulae for $\sin\phi_1$ and $\sin\phi_2$ are valid for all $0\leq\theta < 2\pi$. It remains to determine correctly $\phi_1$ and $\phi_2$, i.e. choosing them from the right quadrants depending on the integration angle $\theta$. Figs.~\ref{fig:circ_phi1} and \ref{fig:circ_phi2} are helpful in this respect. Let us start with $\phi_1$. The phase $\phi_1$ (arrow from $-i$ to $z$) rotates from $0$ to $\pi/2$, when $\theta$ rotates from $0$ to $\cos^{-1}\left (-\tfrac{1}{R}\right )$. Then, when $\theta$ rotates from $\cos^{-1}\left (-\tfrac{1}{R}\right )$ to $\pi$, $\phi_1$ rotates from $\pi/2$ to $\pi$, \textit{etc}. Overall, this yields the following relations
\begin{align}
0\leq\theta <\cos^{-1}\left (-\tfrac{1}{R}\right )\hspace{2em}  &:\hspace{2em} 0\leq\phi_1 <\tfrac{\pi}{2} \nonumber \\
\cos^{-1}\left (-\tfrac{1}{R}\right )\leq\theta <\pi  \hspace{2em}&:\hspace{2em} \tfrac{\pi}{2}\leq\phi_1 <\pi \nonumber \\
\pi\leq\theta <\pi+\cos^{-1}\left (\tfrac{1}{R}\right )  \hspace{2em}&:\hspace{2em} \pi\leq\phi_1 <\tfrac{3\pi}{2} \nonumber \\
\pi+\cos^{-1}\left (\tfrac{1}{R}\right )\leq\theta <2\pi  \hspace{2em}&:\hspace{2em} \tfrac{3\pi}{2}\leq\phi_1 <2\pi .
\end{align}
We can repeat the procedure for $\phi_2$, which regards the arrow drawn from $+i$ to $z$. We get for $\phi_2$
\begin{align}
0\leq\theta <\cos^{-1}\left (\tfrac{1}{R}\right )  \hspace{2em}&:\hspace{2em} \pi\leq\phi_2 <\tfrac{3\pi}{2} \nonumber \\
\cos^{-1}\left (\tfrac{1}{R}\right )\leq\theta <\pi  \hspace{2em}&:\hspace{2em} \tfrac{3\pi}{2}\leq\phi_2 <2\pi \nonumber \\
\pi\leq\theta <\pi+\cos^{-1}\left (-\tfrac{1}{R}\right )  \hspace{2em}&:\hspace{2em} 2\pi\leq\phi_2 <\tfrac{5\pi}{2} \nonumber \\
\pi+\cos^{-1}\left (-\tfrac{1}{R}\right )\leq\theta <2\pi  \hspace{2em}&:\hspace{2em} \tfrac{5\pi}{2}\leq\phi_2 <3\pi .
\end{align}

In general, when evaluating the integral, we can simply use a trapezoidal rule, dividing the $\theta$ interval evenly. While the result is independent of the radius $R$, the appearance of $z\xi$ in the exponent of the exponential function produces accuracy issues for large $R$ and $\xi$---this means, for accurate numerical work, we should use as small an $R$ as possible (we found $R=1.1$ to be good with 100\,000 steps). It also means at some point, the direct numerical integration will fail when $\xi$ is large enough, due to precision issues similar to using the power series to compute $e^{-x}$ for large $x$. 

\begin{figure}
    \centering
    \subfloat{\includegraphics[width=0.45\columnwidth]{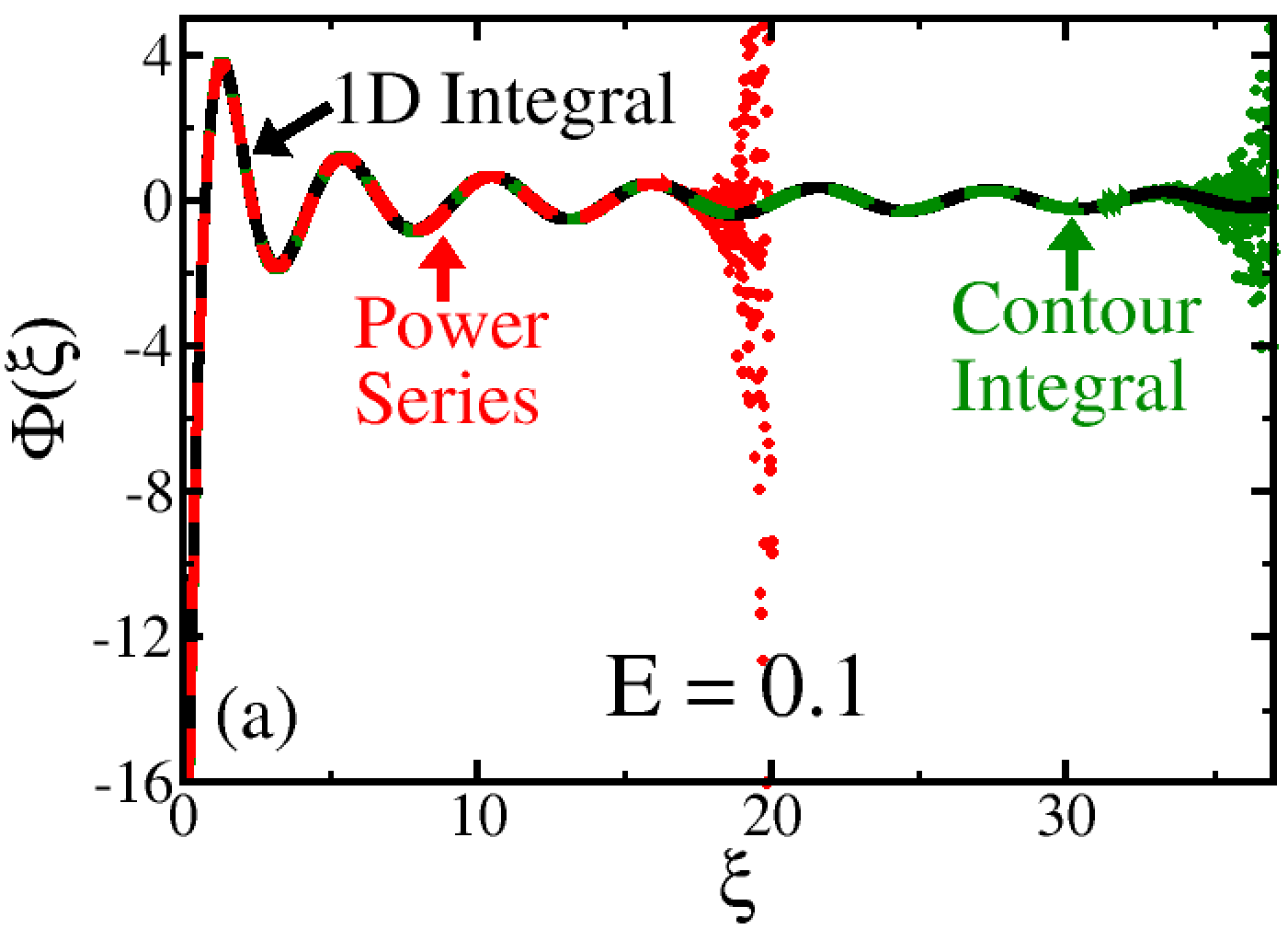}}
    \qquad
    \subfloat{{\includegraphics[width=0.45\columnwidth]{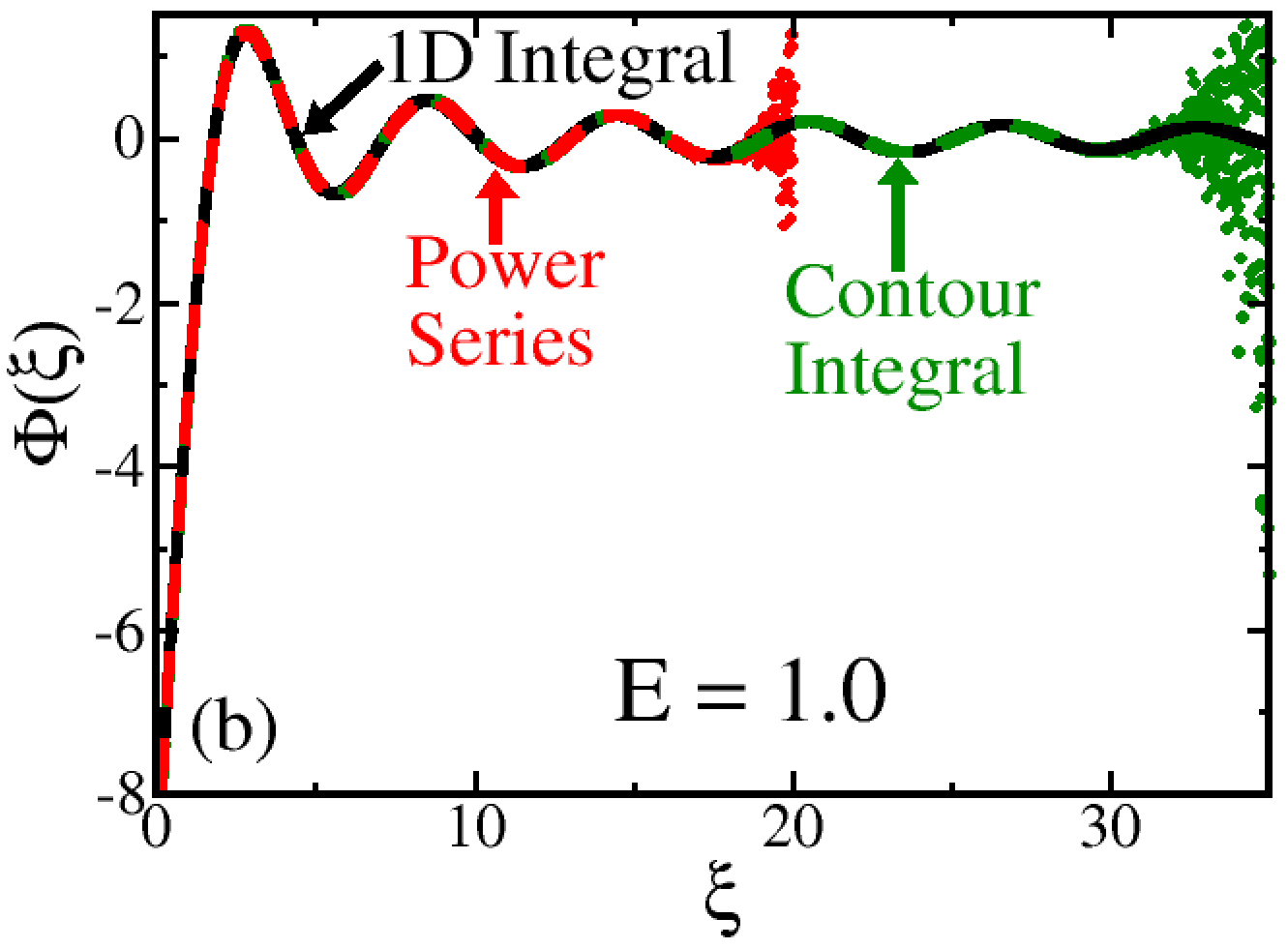} }}
    \\ \subfloat{{\includegraphics[width=0.45\columnwidth]{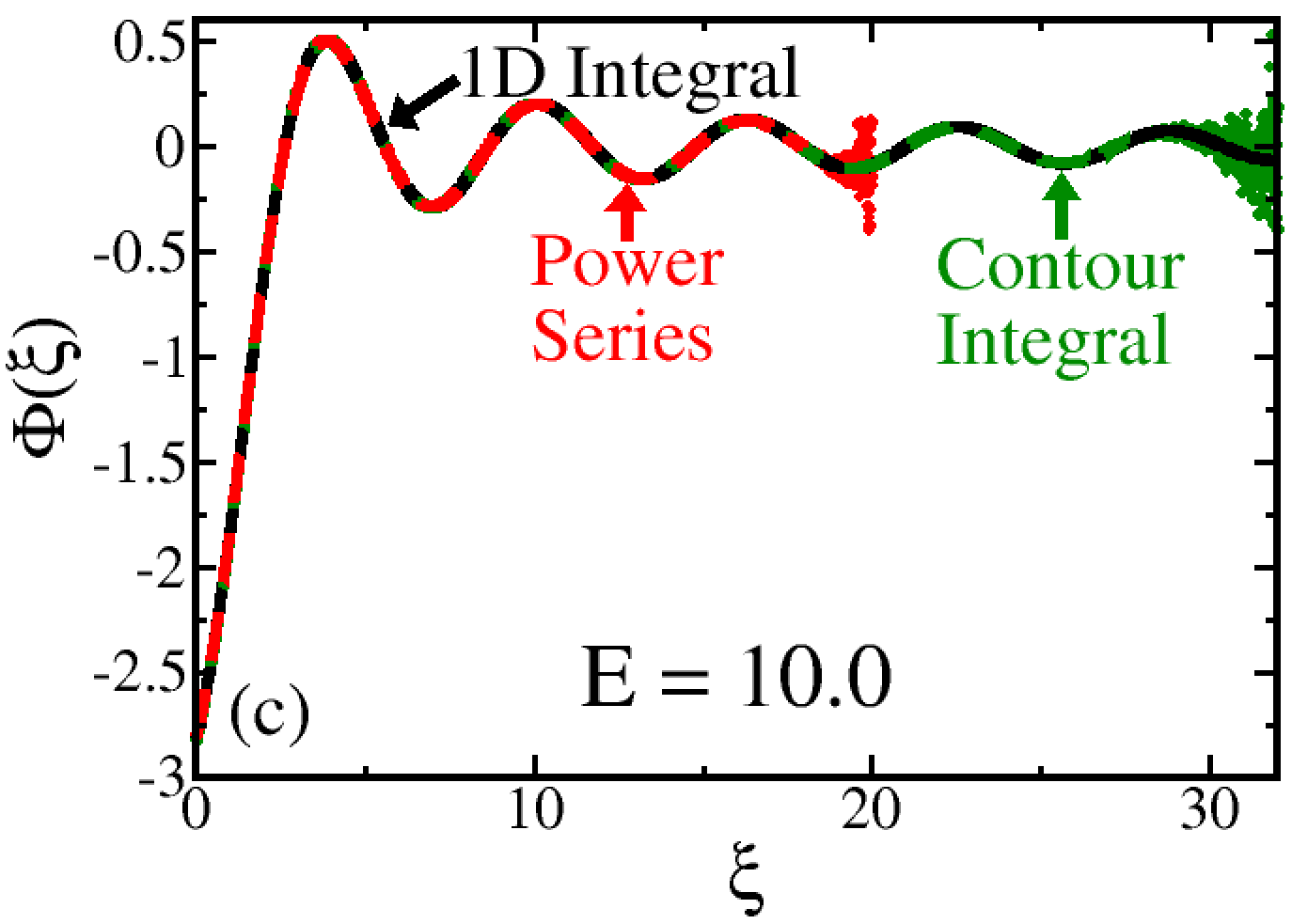} }}
    \caption{Plot of the continuum Coulomb wavefunctions for $l=0$ and three different values of $E$: (a) $E=0.1$; (b) $E=1$, and (c) $E=10$ (all energies are in Hartrees). The power series is shown in red, the contour integral with $R=1.1$ in green, and the one-dimensional integral in black. The three approximations lie on top of each other until they start to fail---the power series fails around $\xi\approx 20$, while the contour integral fails around $\xi\approx 30$. The errors typically occur due to loss of digits of precision in the expressions being evaluated.}
    \label{fig:coulomb_numerics}
\end{figure}

As an example, we plot three results for the radial Coulomb wave function for small $\xi$ and three different energies in the continuum in Fig.~\ref{fig:coulomb_numerics}. These results are normalized according to the result from the 1 dimensional integral for comparison. The results agree with the exact results, as expressed in terms of the confluent hypergeometric function, but requires no knowledge of that function. It instead requires just a moderate computing exercise. We feel that working with these numerical representations of continuum wavefunctions, as derived from the Laplace method, may reduce the cognitive load associated with continuum eigenstates and confluent hypergeometric functions.

All the other continuum solutions except for the Morse potential, proceed in a similar fashion and can be evaluated with the circular contour integral. The Morse potential is different for two reasons: (i) the contour is not given as a rotated dog-bone contour about the imaginary axis and (ii) the function must vanish as $x\to-\infty$. We treat its continuum solutions next.

\section{Continuum solutions of the Morse potential}

The Morse potential differs from the other continuum cases. We see in Table~\ref{Tab:table5} that the Laplace form of the Schr\"odinger equation is once again of the form of Eq.~(\ref{eq:laguerre_diffeq_form}), so $\pm\lambda\in\mathbb{R}$ once again, but now $\beta\in\mathbb{C}$. This differs from both the previous problems of the form in Eq.~(\ref{eq:laguerre_diffeq_form}), where $\beta$ was real, as well as the previous continuum problems, where $\beta$ was an integer. As a result, the solution for $\Phi$ can still be written in the contour integral form given in Eq.~(\ref{eq:gen_laguerre_diffeq}), but the sum of the exponents $\alpha_{+}+\alpha_--2=\beta-2$ is not an integer. This means the integrand is not single-valued as $|z|\to\infty$, so we must draw the branch cut as two pieces, each running from infinity to one of the branch points $\pm\lambda$.  Hence, we draw the branch cut along the real axis in two pieces: one piece travels from $-\lambda$ to $\text{Re}(z)\to-\infty$ and the other piece travels from $\text{Re}(z) = \infty$ to $\lambda$. This branch-cut structure does not allow us to enclose the branch points at $\pm \lambda$ with a contour (as we did for previous continuum cases), nor enclose just one of the two branch points (as in the bound-state problems). However, we can now try another possible contour which connects the two points $\pm\lambda$. For concreteness, we choose it to run through the origin, and we already see a parallel to the other continuum cases, where we took the limiting behavior of a contour running just next to the imaginary axis. For the Morse potential, we have $0\le\xi<\infty$.  Recall that $\xi = 2\frac{\sqrt{2\mu V_0}}{a\hbar} e^{-ax}$, so as $x\to - \infty$ we have $\xi\to\infty$ and as $x\to\infty$ we have that $\xi\to 0$. Since the Morse potential becomes large and positive for $x\to-\infty$, we must have $\psi(x)\to0$. Hence, $\phi(\xi)$ must go to \textit{zero} as $\xi\to \infty$. This is a more stringent condition than just having the wavefunction be bounded as we used previously. This condition eliminates the new contour from $-\lambda\to +\lambda$, because an analysis similar to the previous cases shows that this solution diverges as $\xi\to\infty$. 

Now, let us instead consider the contour as shown in Fig.~\ref{fig:cont_all}b: the contour running from $-\lambda\to - \infty$. The integral solution over that contour is
\begin{equation}
    \Phi(\xi) = \int_{-\lambda}^{-\infty}dz~e^{\xi z}\left(z-\lambda\right)^{\alpha_+-1}\left(z+\lambda\right)^{\alpha_--1},
\end{equation}
where $\lambda=\tfrac{1}{2}$. Now, we make the substitution $z = -t-\lambda$ to obtain 
\begin{equation}
    \Phi(\xi) = e^{-\tfrac{\xi}{2}} (-1)^{\beta-1} \int_0^\infty dt~e^{-\xi t} \left(t+1\right)^{\alpha_+ - 1}t^{\alpha_- -1}.
\end{equation}
Immediately, we see that this integral will go to zero as $\xi\to\infty$ because the exponential term in the integrand guarantees a convergent integral that will vanish in the limit. Moreover, this integral closely resembles the integral representation of the Tricomi confluent hypergeometric function, given by~(Eq.~13.4.4 in Ref.~\cite{dlmf})
\begin{equation}
    U(a,b,z) = \frac{1}{\Gamma(a)}\int_0^\infty dt~e^{-zt}\left(t+1\right)^{b-a-1}t^{a-1}.
\end{equation}
This means we can write the integral solution to the Morse differential equation as
\begin{equation}
    \Phi(\xi) = (-1)^{\beta-1} \Gamma(\alpha_-) e^{-\xi/2} U(\alpha_-,\beta,\xi).
\end{equation}
One might be concerned that this function is not bounded for $\xi\to 0$. Indeed, a simple power-counting argument shows that the magnitude of the  integrand behaves like $\tfrac{1}{z}$ when $\xi=0$. But, because the exponent is complex, it will produce oscillations, which can allow the integral to converge and be bounded. To settle this question, we look at the well-known asymptotics of the Tricomi function $U$. In the limit where the argument $\xi$ goes to zero, the behavior of $U$ is governed by the value of the real part of the second parameter, in our case $\beta$. For the Morse potential, $\text{Re}(\beta) =1 $, so the asymptotic behavior of $U$ as $\xi\to 0$ is (Eq.~13.2.18 of Ref.~\cite{dlmf})
\begin{equation}
    U(\alpha_-,\beta,\xi) = \frac{\Gamma(\beta-1)}{\Gamma(\alpha_-)}\xi^{1-\beta} + \frac{\Gamma(1-\beta)}{\Gamma(1-\alpha_+)} + O(\xi^{2-\text{Re}(\beta)}).
\end{equation}
Once again, $\text{Re}(\beta)=1$, so the $\xi^{1-\beta}$ term will be $\xi^{-i\text{Im}(\beta)}$, which has a modulus of 1 ($\xi^{-i\text{Im}(\beta)}\xi^{i\text{Im}(\beta)}=1$) and a phase that varies rapidly as $\xi\to 0$. Thus, the Tricomi function does not diverge, but oscillates, as $\xi\to 0$. We will see below that the behavior generally looks like a cosine of $x$ for $x\to\infty$.

\begin{figure}
    \centering
    \subfloat{\includegraphics[width=0.45\columnwidth]{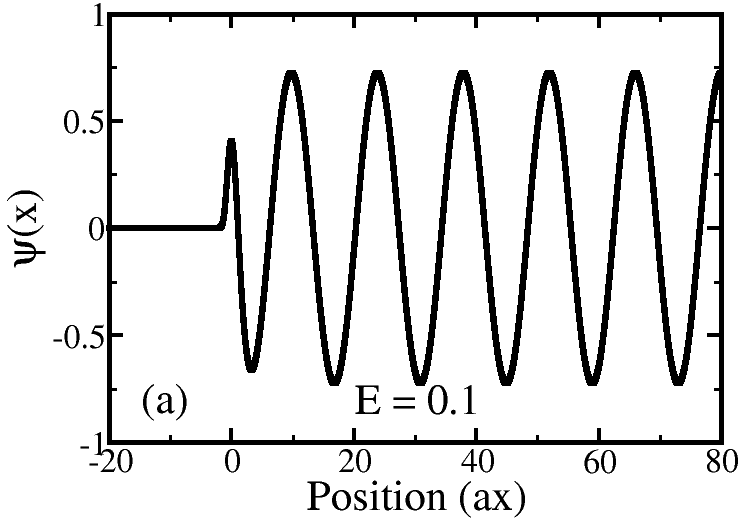}}
    \qquad
    \subfloat{{\includegraphics[width=0.45\columnwidth]{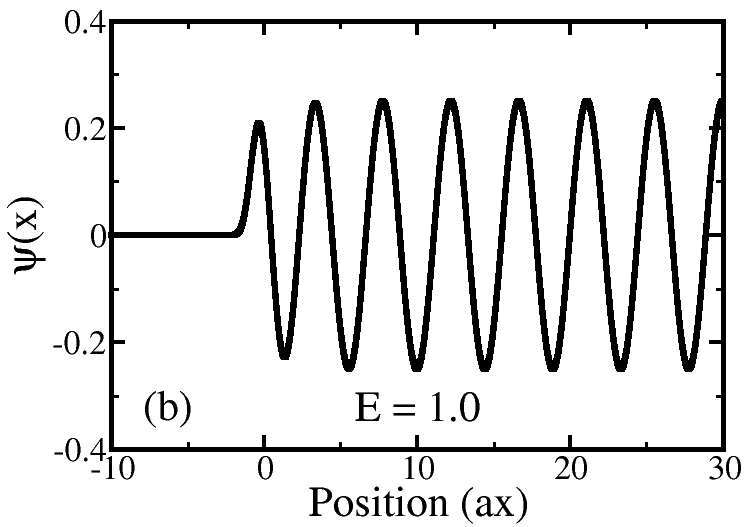} }}
    \\ \subfloat{{\includegraphics[width=0.45\columnwidth]{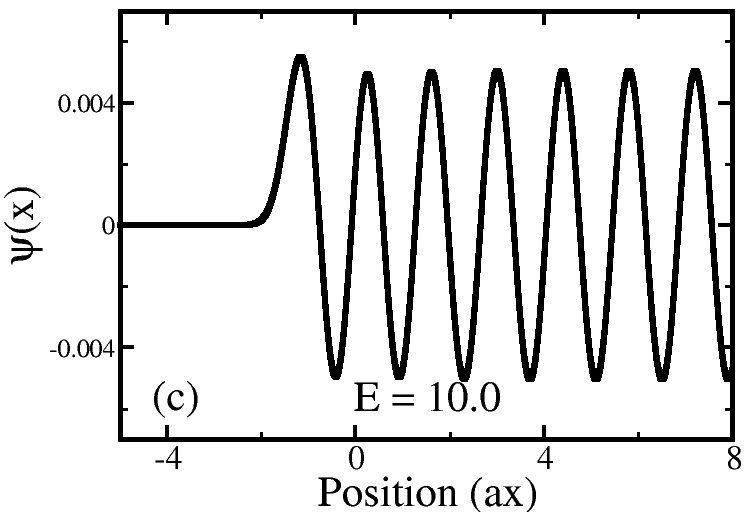} }}
    \caption{Plot of the continuum Morse wavefunction for three different values of $E$: (a) $E=0.1$; (b) $E=1$, and (c) $E=10$. We use $\tfrac{\hbar^2a^2}{2\mu}$ as the energy unit and $\tfrac{1}{a}$ as the length unit. The case we consider is for $V_0=\tfrac{\hbar^2a^2}{2\mu}$. Note how the continuum wave rapidly decays for $x<0$, where the Morse potential becomes large and positive. Because the Morse potential decays to zero exponentially fast, the continuum solution rapidly looks like a simple cosine wave for large positive $x$ with a fixed amplitude. In the region around $x=0$ we see a transition between the two behaviors. }
    \label{fig:morse_wavefunc}
\end{figure}

This contour yields a solution which will be finite everywhere, as well as going to zero as $\xi\to\infty$. Thus, it satisfies all of our requirements, and we can write the solution for the unnormalized Morse potential wavefunction in one dimension as 
\begin{equation}
    \psi(\xi) = \xi^{\frac{\beta-1}{2}}e^{-\frac{\xi}{2}}U(\alpha_-,\beta,\xi),
\end{equation}
Recall that $\xi=2\tfrac{\sqrt{2\mu V_0}}{a\hbar} e^{-ax}$ in this solution. Finally, we note that we can use the so-called Kummer relation (see Eq.~13.2.42 of Ref.~~\cite{dlmf})
\begin{equation}
    U(a,b,z) = \frac{\Gamma(1-b)}{\Gamma(a-b+1)}M(a,b,z)+\frac{\Gamma(b-1)}{\Gamma(a)}z^{1-b}M(a-b+1,2-b,z),
\end{equation}
to relate the above solution in terms of the Tricomi function $U$ to the sum of two complex conjugate Kummer functions $M$, which is the form of the Morse continuum wavefunction that appears in the  literature~\cite{nicholls,matsumoto}. The divergence in each $M$ as $x\to-\infty$ is exactly cancelled by their sum leading to a finite result. We plot the continuum wavefunctions for some typical values in Fig.~\ref{fig:morse_wavefunc}. You can see the behavior is as anticipated. We have a rapid decay for $x<0$, there is a transition region near $x=1$, and then the form is of a constant amplitude sinusoidal oscillation as $x$ increases in the positive direction.

We summarize all of our continuum solution results in two tables. Table~\ref{Tab:table6} shows the results expressed as functions of $\xi$, while Table~\ref{Tab:table7} shows the results in terms of the original variables. The normalization of these wavefunctions is subtle and depends on the scheme that will be used, so we do not discuss the issue of  normalization here.

\begin{table}[ht!]

\centering
 \begin{tabular}{|c| c|} 
 \hline
 \multirow{2}*{\textbf{Problem}} & \textbf{Confluent Hypergeometric}  \\  &\textbf{Form of $\Phi(\xi)$} \\  \hline

 2D Free &  \multirow{2}*{$e^{-i\xi} M\left(|m|+\frac{1}{2},2|m|+1,2i\xi\right)$}  \\ Particle &\\ \hline
 
 3D Free &    \multirow{2}*{$e^{-i\xi}M\left(l+1,2l+2,2i\xi\right)$} \\ Particle &\\ \hline
 
 2D &  \multirow{2}*{$e^{-i\xi}M\left(|m|+\frac{1}{2}+\frac{i\hbar}{a_0\sqrt{2\mu E}},~2|m|+1,~2i\xi\right) $} \\ Coulomb &\\ \hline
 
 3D &   \multirow{2}*{$e^{-i\xi}M\left(l+1+\frac{i\hbar}{a_0 \sqrt{2\mu E}},~2l+2,~ 2i\xi\right) $}  \\ Coulomb &\\ \hline
 
 Morse & \multirow{2}*{$(-1)^{\beta-1} \Gamma(\alpha_-) e^{-\xi/2} U(\alpha_-,\beta,\xi)$}  \\ Potential&  \\
 \hline

 \end{tabular}
 
\caption{Summary of the results of the Laplace method for continuum cases in terms of the variable $\xi$. The solution for $\Phi(\xi)$, as defined in Table~\ref{Tab:table5}, is expressed in terms of the confluent hypergeometric functions $M(a,b,z)$ and $U(a,b,z)$.} 
\label{Tab:table6}

\end{table}

\begin{table}[ht!]

\centering
\resizebox{\textwidth}{!}{
 \begin{tabular}{|c| c|} 
 \hline 
 \textbf{Problem} & \textbf{Unnormalized Wavefunction}\\  \hline

 2D Free &  \multirow{2}*{$J_{|m|}\left(\sqrt{\frac{2\mu E}{\hbar^2}}\rho\right)e^{i|m|\phi}$}\\ Particle &\\ \hline
 
 3D Free &   \multirow{2}*{$j_{l}\left(\sqrt{\frac{2\mu E}{\hbar^2}}r\right)Y_l^m\left(\theta,\phi\right)$} \\ Particle &\\ \hline
 
 2D &  \multirow{2}*{$\rho^{|m|}e^{-i\sqrt{\frac{2\mu E}{\hbar^2}}\rho}e^{i|m|\phi}M\left(|m|+\frac{1}{2}+\frac{i\hbar}{a_0\sqrt{2\mu E}},2|m|+1,2i\sqrt{\frac{2\mu E}{\hbar^2}}\rho\right)$}\\ Coulomb &\\ \hline
 
 3D & \multirow{2}*{$r^{l}e^{-i\sqrt{\frac{2\mu E}{\hbar^2}}r}Y_l^m(\theta,\phi)M\left(l+1+\frac{i\hbar}{a_0 \sqrt{2\mu E}},~2l+2,~ 2i\sqrt{\frac{2\mu E}{\hbar^2}}r\right)$ }\\ Coulomb &\\ \hline
 
 Morse & 
 \multirow{2}*{$\left(\frac{2\sqrt{2\mu V_0}}{a\hbar}e^{-a x}\right)^{i\frac{ \sqrt{2\mu E}}{a\hbar}}\exp\left(-\frac{\sqrt{2\mu V_0}}{a\hbar}e^{-a x}\right)U\left(\frac{ i\sqrt{2\mu E}-\sqrt{2\mu V_0}}{a\hbar}+\frac{1}{2},2i\frac{ \sqrt{2\mu E}}{a\hbar}+1,\frac{2\sqrt{2\mu V_0}}{a\hbar}e^{-a x}\right)$}\\ Potential &   \\ 
 \hline

 \end{tabular}
 }
\caption{Summary of the results of the continuum cases we solved with the Laplace method in terms of the original independent variable.  For the free particle cases, we express the more common form of the confluent hypergeometric function.
} 
\label{Tab:table7}

\end{table}


\section{Pedagogical discussion}

Back in 1937, Dirac urged quantum instruction to include treatments via complex analysis~\cite{dirac}. This suggestion appears to have been ignored by all (including Dirac himself) as it never made it into later editions of his quantum mechanics textbook. This appears to us to have been an unfortunate mistake. As one studies more advanced quantum field theory for particle physics, or many-body physics for condensed matter, facility with complex analysis greatly helps in making progress with the material. So, an earlier quantum mechanics class would be an ideal setting to start developing facility with complex analysis ideas within quantum instruction.

Furthermore, with the emphasis on the series solution (Fr\"obenius method) for solving bound state problems, one does not develop the tools to tackle the continuum solutions. Hence, in advanced quantum classes, where these are discussed, they usually are covered in a somewhat \textit{ad hoc} fashion that emphasizes analytic continuation of the bound-state solutions and then the student being simply told what the answer is. In our opinion, this is \textit{not} teaching the material. Instead, the approach we present here, via the Laplace method (building off of Schr\"odinger's original solution for hydrogen) provides a nice approach to cover continuum solutions and introduces a proper way to teach complex analysis ideas within a graduate quantum mechanics setting. The integration over the circle of fixed radius provides a nice way to determine these functions numerically, using a relatively simple code, once one knows how to properly determine the phases of the powers given the branch cuts used.

It is true that one can instead be told the continuum solution and substitute it into the Schr\"odinger equation and, by employing identities of the confluent hypergeometric functions, verify that it solves the differential equation. But, this is not actually done in advanced quantum textbooks, nor is it commonly done in quantum instruction, probably because it is deemed to be too much work to have students find and properly use the required confluent hypergeometric equation identities (even if they are straightforward to use). In this circumstance, the approach presented here provides a useful alternative to other methodologies.

How do we work with these ideas with students? We think that one can present the solution for some example problems, such as hydrogen, in class and then ask the students to produce results for other potentials on  homework exercises (except the Morse potential, of course, because that problem requires different methods in the continuum and is a bit more subtle in its analysis). In addition to developing a capacity for working with these solutions, we strongly believe students should numerically calculate these functions using one of the integral forms (we prefer the integration over a circle of fixed radius for the continuum problems), in order to develop numerical skills within a quantum context and to learn the challenges with accurately determining these functions for large arguments.

\section{Conclusions}

We have shown how the almost forgotten Laplace method can be employed as a powerful tool within graduate quantum instruction to teach how to determine bound and continuum states of all problems that are solved with confluent hypergeometric equation wavefunctions. The approach helps include complex analysis instruction within a quantum setting, which we believe will better prepare students for more advanced quantum field theory in a high-energy or condensed-matter setting. The bound-state solutions provide a way to relate the orthogonal polynomials that arise to their Rodrigues formulas, which naturally emerge via the poles in the contour integrals needed to determine the wavefunctions. Quantization of the energies in the bound states also occurs naturally. For the continuum solutions, we find a simple contour integral that gives us the wavefunctions, but it requires a precise determination of the phases of the terms that are raised to complex-valued powers---hence it requires a proper understanding of branch cuts and how to evaluate the polar radius and phase of complex numbers on a cut plane---a quite valuable skill for graduate students to learn.

We believe that using this approach provides students with well-needed practice in the use of complex analysis methods. It also allows for a solution of the linear potential problem (which we did not discuss here). That problem is treated with methods similar to what we developed here in other textbooks, in particular, the textbook by Konishi and Paffuti~\cite{konishi_paffuti} has a nice treatment of this. Curiously, Landau and Liftshitz \cite{landau_lifshitz} use the Laplace method to determine the properties of the confluent hypergeometric functions in their appendix, but they do not use the Laplace method for finding solutions to the Schr\"odinger equation in the main part of the text.

The materials we present here are most appropriate for graduate quantum instruction and, if used, help differentiate the graduate class from the undergraduate one, rather than just repeating the undergraduate one with harder problems, as is commonly done.

\end{document}